\documentclass{aa}  
\usepackage{graphicx}
\usepackage{booktabs}
\usepackage{supertabular}
\usepackage[version=4]{mhchem}
\usepackage{ulem}
\usepackage{xcolor}
\usepackage{amsmath}
\usepackage{sidecap}
\usepackage{caption}
\usepackage{float}
\usepackage{txfonts}
\usepackage{hyperref}
\hypersetup{
    colorlinks=true,
    linkcolor= blue,
    filecolor=magenta,      
    urlcolor=blue,
    citecolor= blue,
}%
\begin{document}

   \title{Correlations among complex organic molecules around protostars: Effects of physical structure}


   \author{P. Nazari,
          \inst{1}
          \and
          B. Tabone\inst{2}
          \and
          G. P. Rosotti\inst{1,3}
          \and
          E. F. van Dishoeck\inst{1,4}
          }

   \institute{Leiden Observatory, Leiden University, P.O. Box 9513, 2300 RA Leiden, the Netherlands\\ 
        \email{nazari@strw.leidenuniv.nl}
         \and
        Universit\'e Paris-Saclay, CNRS, Institut d’Astrophysique Spatiale, 91405 Orsay, France
        \and
        Dipartimento di Fisica `Aldo Pontremoli', Universit\`{a} degli Studi di Milano, via G. Celoria 16, I-20133 Milano, Italy
        \and
        Max Planck Institut f\"{u}r Extraterrestrische Physik (MPE), Giessenbachstrasse 1, 85748 Garching, Germany
             }

   \date{Received 29 May 2023 / Accepted 14 April 2024}

 
  \abstract
   {Complex organic molecules have been observed toward many protostars. Their column density ratios are generally constant across protostellar systems with some low-level scatter. However, the scatter in formamide (NH$_2$CHO) to methanol (CH$_3$OH) column density ratio, $N_{\rm NH_2CHO}/N_{\rm CH_3OH}$, is one of the highest compared with other ratios. The larger scatter for $N_{\rm NH_2CHO}/N_{\rm CH_3OH}$ (or weak correlation of these two molecules) is sometimes interpreted as evidence of gas-phase formation of NH$_2$CHO.}
   {In this work we propose an alternative interpretation in which this scatter is produced by differences in the snowline locations related to differences in binding energies of these species (formamide typically has ${\gtrsim}2000$\,K larger binding energy than methanol) and the small-scale structure of the envelope and the disk system. Therefore, we do not include chemistry in our models to isolate the effect of physical factors. We also include CH$_3$CN in our work as a control molecule which has a similar binding energy to CH$_3$OH.}
   {We use radiative transfer models to calculate the emission from NH$_2$CHO, CH$_3$OH, and CH$_3$CN in protostellar systems with and without disks. The abundances of these species are parameterized in our models. Then we fit the calculated emission lines to find the column densities and excitation temperatures of these species as done in real observations.}
   {Given the difference in binding energies of NH$_2$CHO and CH$_3$OH, we find a correction factor of ${\sim}10$ to be multiplied by gas-phase $N_{\rm NH_2CHO}/N_{\rm CH_3OH}$ to give the true abundance ratio of these two species in the ices. This factor is much smaller (i.e., ${\sim}2$) for $N_{\rm CH_3CN}/N_{\rm CH_3OH}$ (the control molecule). We find that models with different disk sizes, luminosities and envelope masses produce a scatter in this correction factor and hence in $N_{\rm NH_2CHO}/N_{\rm CH_3OH}$, comparable with that of observations. The scatter in $N_{\rm NH_2CHO}/N_{\rm CH_3OH}$ is larger than that of $N_{\rm CH_3CN}/N_{\rm CH_3OH}$ in models consistent with the observations. However, the scatter in models for $N_{\rm CH_3CN}/N_{\rm CH_3OH}$ is smaller than observations by a factor of ${\sim} 2$ as expected from the similar binding energies of CH$_3$OH and CH$_3$CN pointing to the need for some chemical effects in the gas or ice to explain those observed ratios. We show that the scatter in $N_{\rm NH_2CHO}/N_{\rm CH_3OH}$ will be lower than previously measured if we correct for the difference in sublimation temperatures of these two species in observations of ${\sim} 40$ protostellar systems with ALMA.}
   {The scatter in $N_{\rm NH_2CHO}/N_{\rm CH_3OH}$ (or ratio of any two molecules with large binding energy difference) can be partially explained by the difference in their binding energies. Correction for this bias makes the scatter in this ratio similar to that in ratios of other complex organics in the observations, making NH$_2$CHO a `normal' molecule. Therefore, we conclude that gas-phase chemistry routes for NH$_2$CHO are not necessary to explain the larger scatter of $N_{\rm NH_2CHO}/N_{\rm CH_3OH}$ compared with other ratios.}

   \keywords{Astrochemistry --
                Stars: protostars --
                ISM: abundances --
                ISM: molecules --
                Radiative transfer
               }

   \maketitle

\section{Introduction}

The protostellar phase is one of the richest phases of star formation in species such as complex organic molecules. These species are defined as those having six or more atoms including carbon and hydrogen (\citealt{Herbst2009}; \citealt{Ceccarelli2017}). Numerous studies have focused on analyzing these molecules toward both low- and high-mass protostars (\citealt{Blake1987}; \citealt{Ewine1995}; \citealt{Beltran2009}; \citealt{Jorgensen2016}; \citealt{Rivilla2017}; \citealt{vanGelder2020}; \citealt{McGuire2021}; \citealt{Baek2022}; \citealt{Codella2022}; \citealt{Ligterink2022}). 

Many of such studies look for correlations between column densities of various complex organics as a clue on their formation pathways (e.g., \citealt{Belloche2020}; \citealt{Coletta2020}; \citealt{Ligterink2020}; \citealt{Allen2020}; \citealt{Law2021}; \citealt{Martin2021}; \citealt{Taniguchi2023}). In particular, the lack of correlation between observed column densities (or large scatter in column density ratios) for many sources is often interpreted as importance of gas-phase chemical routes in production/destruction of a molecule (e.g., \citealt{Yang2021}; \citealt{Chahine2022}). This is because a small scatter in the observed column density ratios (i.e., strong correlation) found for various sources indicates similar physical conditions for the formation environment of these species in different sources (\citealt{Quenard2018};  \citealt{Belloche2020}), which is more probable to be achieved on ices in the pre-stellar phase rather than in the gas (\citealt{Coletta2020}). Hence, if large scatters are observed, their formation environments are thought to be different and that is achieved easier in the gas. However, the latter conclusion should be made with caution.

Chemical effects are not the only way of producing a scatter in the observed column density ratios of two species. Other physical effects such as source structure can affect column density ratios of molecules with different binding energies and produce a scatter. This is because in measurement of column densities from spatially unresolved observations it is often assumed that the emitting size is the same for all species around a single protostar. However, if two molecules have different binding energies (i.e., different emitting sizes) this assumption breaks down in the protostellar disk and envelope that have a temperature gradient. 

Recently, \cite{Nazari2022ALMAGAL} analyzed the Nitrogen-bearing complex organics (plus methanol as a reference; \citealt{vanGelder2022}) around ${\sim} 40$ massive protostars and found that in general column density ratios of different pairs are remarkably constant across low- and high-mass protostars with a small degree of scatter (mostly a factor $\lesssim 2.5$ around the mean). They concluded that the constant ratios point to the similarity of the environment in which these species form in, likely pre-stellar ices (also see \citealt{Chen2023}). However, they found a larger scatter in column density ratios (factor ${\sim} 3$ around the mean) for molecules with different binding energies. They speculated that the reason for this large scatter could be the difference between location of sublimation fronts of those molecules which would result in a correction factor to be applied to the column density ratios. However, they explain that this correction factor will be roughly constant for each set of two species and will not result in a scatter unless the protostellar systems have different source structures. In which case the correction factor will be different for each source. In this work we examine how source structure can cause variations in this correction factor and thus column density ratios of formamide (NH$_2$CHO) to methanol (CH$_3$OH), two molecules with different binding energies and ratios that showed one of the largest scatters in previous observations.  

In this work we specifically focus on formamide and methanol because methanol is known to have a relatively low (${\sim} 6500$\,K) binding energy and formamide a high (${\sim} 9500$\,K) binding energy (\citealt{Penteado2017}; \citealt{Wakelam2017}; \citealt{Chaabouni2018}; \citealt{Ferrero2020}; \citealt{Minissale2022}). Therefore, they are expected to trace regions with low temperatures (farther from the protostar) and high temperatures (close to the protostar), respectively. This was also suggested based on the measured excitation temperatures of these two species in \cite{Nazari2022ALMAGAL}, with values for methanol clustering at $T_{\rm ex} \simeq 100$\,K and for formamide at $T_{\rm ex}\simeq 300$\,K. A difference between the emitting areas of formamide and methanol was also observed with spatially resolved observations toward one source (HH212; \citealt{Lee2022}). Given the large difference between their binding energies, the ratio of $N_{\rm NH_2CHO}/N_{\rm CH_3OH}$ is the best combination to study the effect of source structure on the scatter in column density ratios.

Another reason for considering formamide is the ongoing debate regarding its formation route. Although it is generally agreed that methanol forms on the surfaces of grains (\citealt{Watanabe2002}; \citealt{Fuchs2009}), it is less clear whether formamide forms in the gas phase or on solids, with both gas and ice formation pathways having been suggested for formamide (\citealt{Jones2011}; \citealt{Barone2015}; \citealt{Codella2017}; \citealt{Haupa2019}; \citealt{Douglas2022}). Traditionally the large scatter seen in $N_{\rm NH_2CHO}/N_{\rm CH_3OH}$ among various sources would be interpreted as gas-phase formation of formamide. However, if physical effects such as differences in source structure can produce the observed scatter in $N_{\rm NH_2CHO}/N_{\rm CH_3OH}$, gas-phase formation routes are not necessary for formamide.

Significant scatter in column density ratios as a result of source structure variations is expected only for species with a large difference in their binding energies (e.g., NH$_2$CHO/CH$_3$OH). However, methyl cyanide (CH$_3$CN) is also included in this paper as a `control' molecule because methanol and methyl cyanide are expected to have similar binding energies (${\sim} 6500$\,K; \citealt{Minissale2022}). Therefore, one does not expect a scatter in $N_{\rm CH_3CN}/N_{\rm CH_3OH}$ if no chemistry is included in the models. An example of this additional chemistry for CH$_3$CN could be the gas-phase formation routes predicted by some chemical models (\citealt{Garrod2022}; \citealt{Taniguchi2023}) or variations in initial ice abundances due to grain surface chemistry.

In this paper, the envelope-only and envelope-plus-disk models of low- and high-mass protostars are used to calculate the line emissions of methanol, formamide and methyl cyanide using radiative transfer and taking parametrized abundances. Next, these lines are fitted to find the column densities in the same way as normally done in the observational analysis. We then find the column density ratios in models with a range of luminosities, envelope masses and disk sizes. The scatter in the column density ratios are measured and compared with the findings from observations.

\section{Methods}
\label{sec:methods}

\subsection{Radiative transfer models}
\label{sec:rad_models}

\begin{figure*}
    \centering
    \includegraphics[width=\textwidth]{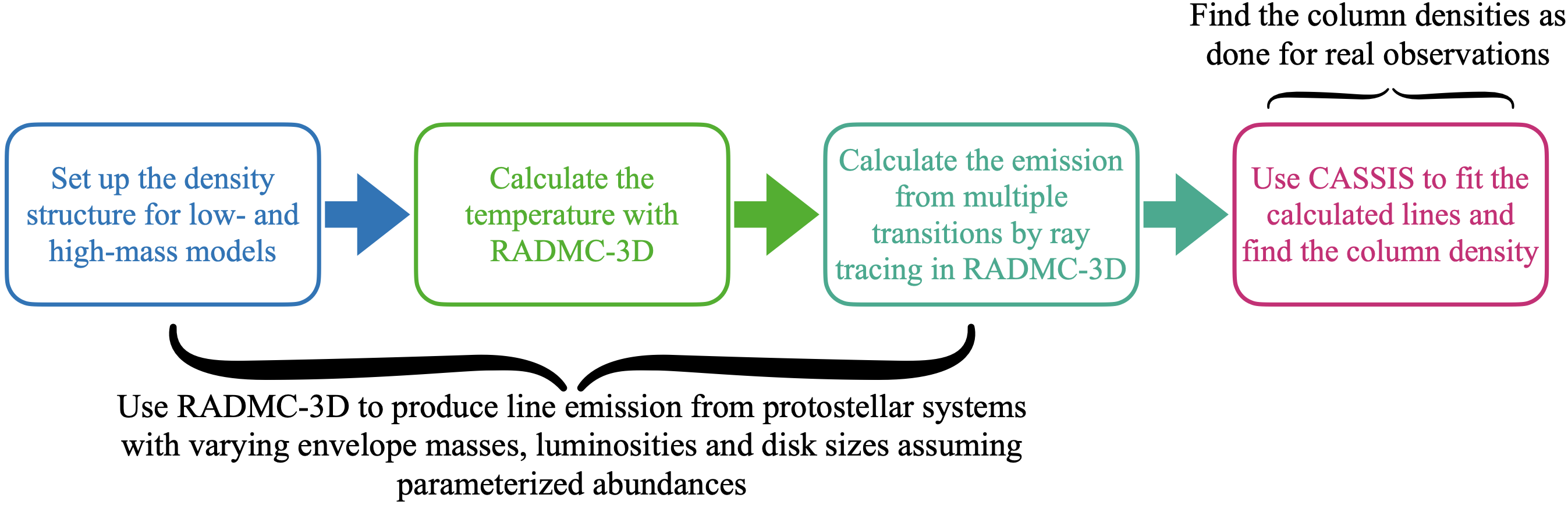}
    \caption{Schematic of our methods. First RADMC-3D is used to produce the line emission from multiple transitions of methanol, formamide, and methyl cyanide for a grid of models. Then we use CASSIS to fit those lines and find the column densities.} 
    \label{fig:flowchart}
\end{figure*}

This work uses the same models studied in \citet{Nazari2022, Nazari2023} to simulate the temperature structure of and line emission from low- and high-mass protostellar systems. A schematic of our methods is presented in Fig. \ref{fig:flowchart}. An envelope-only model and an envelope-plus disk model were considered with the same physical structures as in the two papers mentioned above (see Fig. \ref{fig:density_scatterpaper}). In low-mass protostellar disk models viscous heating was not included but it was taken into account in the high-mass protostellar disk models due to the higher accretion rates of these objects (\citealt{Hosokawa2009}; \citealt{Beuther2017}). For the temperature and line emission calculation the code RADMC-3D (\citealt{Dullemond2012}) version 2.0\footnote{\url{http://www.ita.uni-heidelberg.de/~dullemond/software/radmc-3d}} was used assuming local thermodynamic equilibrium (LTE) excitation. The grid and number of photons used in these calculations were the same as in \citet{Nazari2022, Nazari2023} for low- and high-mass protostars. A sub-sample of all the modeled protostars in those studies were considered here while keeping the range of parameters appropriate for most observations. Table \ref{tab:params} shows the parameters of the models studied in this work.     

\begin{table*}[t]
  \centering
  \caption{Parameters of the models}
  \label{tab:params}
  \resizebox{\textwidth}{!}{\begin{tabular}{@{\extracolsep{1mm}}*{6}{l}}
          \toprule
          \toprule    
          & \multicolumn{2}{c}{Envelope-only}&\multicolumn{2}{c}{Envelope-plus-disk} & \\
          \cmidrule{2-3} \cmidrule{4-5}
                Parameter [unit] & low-mass & high-mass & low-mass & high-mass & Description \\
                \midrule     
  
    $r_{\rm in}$ [au] & 0.4 & 10 & 0.4 & 10 & The inner radius\\
    $r_{\rm out}$ [au]  & $10^{4}$ & $5\times10^{4}$ & $10^{4}$ & $5\times10^{4}$ & The outer radius of the envelope\\
    $M_{\rm E}$ [M$_{\odot}$] & \textbf{1}, 3, 5 & 50, \textbf{300}, 1000 & \textbf{1}, 3, 5 & 50, \textbf{300}, 1000 & Envelope mass\\
    $R_{\rm D}$ [au] & -- & -- & 20, \textbf{50}, 200 & 500, \textbf{1000}, 2000 & Disk radius\\
    $T_{\star}$ [K]  & 5000 & 40000 & 5000 & 40000 & Protostellar temperature\\
    $M_{\star}$[M$_{\odot}$]  & 0.5 & 30 & 0.5 & 30 & Protostellar mass\\
    $L$ [L$_{\odot}$] & 2, \textbf{8}, 32 & 5$\times$10$^2$, \textbf{10$^4$}, 5$\times$10$^5$ & 2, \textbf{8}, 32 & 5$\times$10$^2$, \textbf{10$^4$}, 5$\times$10$^5$ & Bolometric luminosity\\
    \bottomrule
  \end{tabular}}
  \tablefoot{The fiducial model parameters are highlighted in boldface. The fiducial models are those with small dust grains. When calculating the emission lines we assume source distances of 150\,pc and 4\,kpc for low- and high-mass protostars.}
\end{table*}


Here three molecules are considered: CH$_3$OH, NH$_2$CHO and CH$_3$CN. The abundances of these species were parameterized. They were calculated using the balance between adsorption and thermal desorption (\citealt{Hasegawa1992}). The calculation of this balance is slightly different from \cite{Nazari2022, Nazari2023} following the recommendations of \cite{Minissale2022} and \cite{Ligterink2023}. We now include the appropriate pre-factor when calculating the ice and gas abundances. This results in modifying Eq. (6) of \cite{Nazari2022} to

\begin{equation}
    \frac{X_{\rm ice}}{X_{\rm gas}} = \frac{\pi a_{\rm d}^{2} n_{\rm d} S \sqrt{3k_{\rm B}T_{\rm gas}/m_{i}}}{e^{-E_{\rm b}/T_{\rm d}} \nu_{\rm TST}},
    \label{eq:gas-grain_new_prefactor}
\end{equation}

\noindent where $a_{\rm d}$ is the dust grain size, $n_{\rm d}$ is the dust number density, $S=1$ is the sticking coefficient, $E_{\rm b}$ is the binding energy, $m_{\rm i}$ is the mass of the considered species i, and $T_{\rm gas}$ and $T_{\rm d}$ are the gas and dust temperatures, respectively. In Eq. \eqref{eq:gas-grain_new_prefactor} $\nu_{\rm TST}$ is the pre-factor calculated from the transition state theory (TST). Binding energies of 6621\,K, 9561\,K and 6253\,K, and pre-factors of $3.18 \times 10^{17}$, $3.69 \times 10^{18}$, and $2.37 \times 10^{17}$ are assumed for methanol, formamide and methyl cyanide, respectively, based on the recommended values by \citealt{Minissale2022}. The difference between sublimation temperature of methanol in this work and \cite{Nazari2022} is around 35\,K for a typical envelope density. This difference results in a factor of ${\sim}5$ difference between the line fluxes. More discussion on how variation of binding energy changes the column density ratios is given in Sect. \ref{sec:column_ratios}. To keep the line fluxes between this work and \cite{Nazari2022, Nazari2023} consistent we increase the abundance of methanol in the disk and envelope by a factor of 5. This is justified given the range of methanol ice abundances observed in \cite{Oberg2011} and \cite{Boogert2015}. Total CH$_3$OH gas and ice abundances ($X_{\rm gas}+X_{\rm ice}$) of $5\times10^{-6}$ and $5\times10^{-9}$ in the envelope and the disk are assumed with minimum $X_{\rm gas}$ of $5 \times 10^{-9}$ and $5 \times 10^{-11}$ outside the snow lines in the envelope and the disk. Here we model the line emission from the $^{18}$O isotopologue of CH$_3$OH assuming $^{16}$O/$^{18}$O of 560 and 400 for low- and high-mass protostars from observations (\citealt{Wilson1994}; \citealt{vanGelder2020}) to avoid optically thick lines. The abundances of NH$_2$CHO and CH$_3$CN in the envelope and the disk were scaled from the CH$_3$OH abundances using the observed gaseous column density ratios of NH$_2$CHO/CH$_3$OH (${\sim}2 \times 10^{-3}$) and CH$_3$CN/CH$_3$OH (${\sim}8 \times 10^{-3}$) from \cite{Nazari2022ALMAGAL} who found similar mean values for column density ratios between low- and high-mass protostars. Hence, we multiplied the above abundances of CH$_3$OH in the models by $2 \times 10^{-3}$ and $8 \times 10^{-3}$ to parameterize the abundances of formamide and methyl cyanide. To avoid optically thick CH$_3$CN lines we adopted the same strategy as used for methanol by calculating the $^{13}$CH$_3$CN emission lines. We divided the abundances found from column density ratios of CH$_3$CN/CH$_3$OH by the $^{12}$C/$^{13}$C isotopologue ratio of 70 in our low- and high-mass protostars (\citealt{Milam2005}; \citealt{Jorgensen2016}; \citealt{vanGelder2020}). We note that although the abundances for each molecule are chosen based on chemical models and observations (\citealt{Walsh2014}; \citealt{Boogert2015}; \citealt{Nazari2022ALMAGAL}), the exact abundances will not change the conclusions of this work because the scatter is the main interest here.

\begin{figure*}
    \centering
    \includegraphics[width=0.8\textwidth]{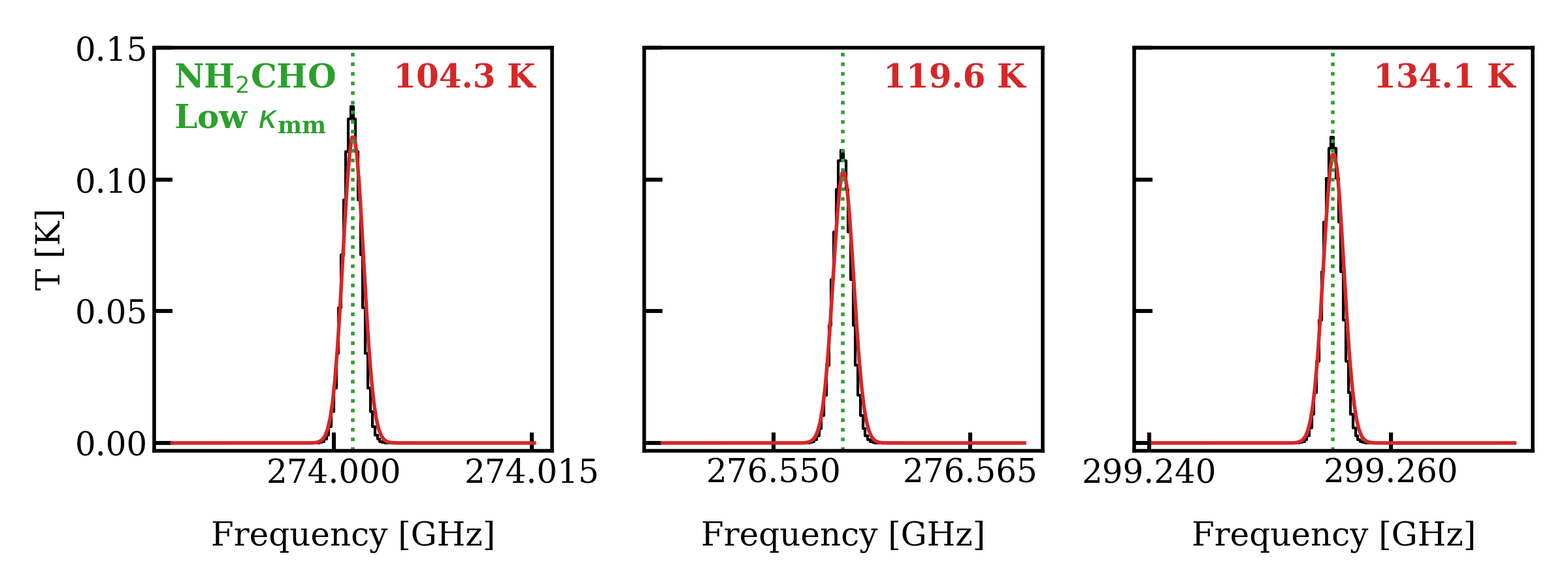}
    \caption{Simulated data from RADMC-3D in black with the CASSIS model fitted in red with $T_{\rm ex}$ fixed to 150\,K. The upper energy levels are printed on the top right of each panel. The dotted green lines mark the transition frequency of each line. This is for the fiducial envelope-only low-mass model ($L=8$\,L$_{\odot}$, $M_{\rm E}=1$\,M$_{\odot}$ with low mm opacity dust). See Figures \ref{fig:fit_fid_large}-\ref{fig:fit_fid_large_CH3OH} for other fitting examples.} 
    \label{fig:fit_fid}
\end{figure*}

We emphasize that an important assumption in this work is that the species are primarily formed on the grains and are sublimated into the gas close to the protostar without any further chemical reactions. In other words, the abundances remain constant in the gas phase and do not change as a function of temperature. This assumption is necessary to isolate the effect of physical structure on the scatter in column density ratios. There are already works studying the effects of chemistry on the correlation between species (e.g., \citealt{Quenard2018}; \citealt{Skouteris2018}; \citealt{Taniguchi2023}) but that is not the goal here. In other words, we want to examine how much of the (anti-)correlation among molecules is purely due to variations in physical structure.

Ray tracing was done in the same way and with similar dust distributions to \citet{Nazari2022, Nazari2023} but for more lines in this work. Here we consider multiple CH$_3$OH, NH$_2$CHO and CH$_3$CN lines for subsequent column density measurement. The line data for the three molecules were taken from the Leiden Atomic and Molecular Database (\citealt{Schoier2005}). Each line was calculated assuming that the low-mass protostars were located at 150\,pc and the high-mass ones at 4\,kpc. Two dust distributions were considered; namely low millimeter (mm) and high mm opacity dust grains which are representative of small and large dust grain distributions. The dust opacities for these two cases at a wavelength of 1\,mm are ${\sim}$0.2\,cm$^2$\,g$^{-1}$ and ${\sim}$18\,cm$^2$\,g$^{-1}$, respectively. The line emissions were initially calculated including dust grains in ray tracing but subsequently the lines were continuum subtracted. The studied lines are presented in Table \ref{tab:lines}. The lines used for most models were chosen such that all have an upper energy ($E_{\rm up}$) of between 100\,K and 200\,K because those lines trace the bulk of the gas inside of the sublimation fronts. To investigate the implications and effects of this assumption, we include lines with a range of $E_{\rm up}$ (highlighted with stars in Table \ref{tab:lines}) for the fiducial low- and high-mass models for excitation temperature analysis. Moreover, the frequency of the lines were chosen to be mainly in Band 6 and 7 of the Atacama Large Millimeter/submillimeter Array (ALMA) similar to the observations that will be compared with these models. Given that the spectral fitting process assumes LTE conditions, here we did the ray tracing assuming LTE. In this work we are mainly concerned with the ratio of column densities and the emitting areas especially close to the protostar and hence the assumption of LTE should not change the conclusions (see also discussion on non-LTE conditions in \citealt{Nazari2022}).

\subsection{Measurement of $N$ and $T_{\rm ex}$ from models}
\label{sec:col_calc}

Once the radiative transfer models produce the line emissions, we fit those lines to obtain the column densities in the same way as done in observational analysis (Fig. \ref{fig:flowchart}). The lines with $E_{\rm up}$ between 100\,K and 200\,K were fitted together using the spectral analysis tool CASSIS \footnote{\url{http://cassis.irap.omp.eu/}} (\citealt{Vastel2015}) assuming LTE. The line lists were taken from the Cologne Database for Molecular Spectroscopy (CDMS; \citealt{Kukolich1971}; \citealt{Moskienko1991} ; \citealt{Muller2001}; \citealt{Muller2005}; \citealt{Cazzoli2006}; \citealt{Xu2008}).

We fit the lines for each source using a grid fitting method also used in \cite{vanGelder2020}. We made a grid of column densities between 10$^{9}$\,cm$^{-2}$ and 10$^{17}$\,cm$^{-2}$ on logarithmic scales with 0.1 spacing. We also fitted for the full width at half maximum (FWHM) in a grid with resolution of 0.4\,km\,s$^{-1}$ and ranges of 0.2-2\,km\,s$^{-1}$ and 0.2-4\,km\,s$^{-1}$ for low- and high-mass protostars, respectively. The final models normally have FWHM of 2\,km\,s$^{-1}$ and 4\,km\,s$^{-1}$ for low- and high-mass protostars but for a few models (especially some of those with disks) the intensity at the line peak becomes narrow and is best to fit for FWHM. Moreover, we fixed the excitation temperature to 150\,K when finding the column densities from the lines with $E_{\rm up}$ between 100\,K and 200\,K. Fixing the temperature anywhere between 100\,K and 200\,K does not change the column densities significantly (within a factor 2) as found also in other works (e.g., \citealt{Taquet2015}; \citealt{Coutens2016}; \citealt{Ligterink2021}; \citealt{vanGelder2022deuteration, vanGelder2022}; \citealt{Chen2023}). In all cases a single-component CASSIS model is fitted. In this procedure, first column densities of $^{18}$O and $^{13}$C isotopologues of methanol and methyl cyanide are calculated and then the column densities are multiplied by the respecting isotopologue ratios mentioned in Sect. \ref{sec:rad_models}.

An uncertainty of 20$\%$ was assumed on column densities to take into account the calibration error in real observations. In the calculation of the lines a global beam size of $2\arcsec$ was used to mimic the angular resolution of observations. Moreover, similar to observational studies a beam dilution of 1 (i.e., no beam dilution) was assumed when calculating the column densities in CASSIS. Given that only the ratio of column densities is of interest, the assumption for beam dilution is not important as long as the lines stay optically thin. Figures \ref{fig:fit_fid} and \ref{fig:fit_fid_large}-\ref{fig:fit_fid_large_CH3OH} present examples of line fitting for NH$_2$CHO and CH$_3^{18}$OH. 

\begin{figure*}
    \centering
    \includegraphics[width=0.8\textwidth]{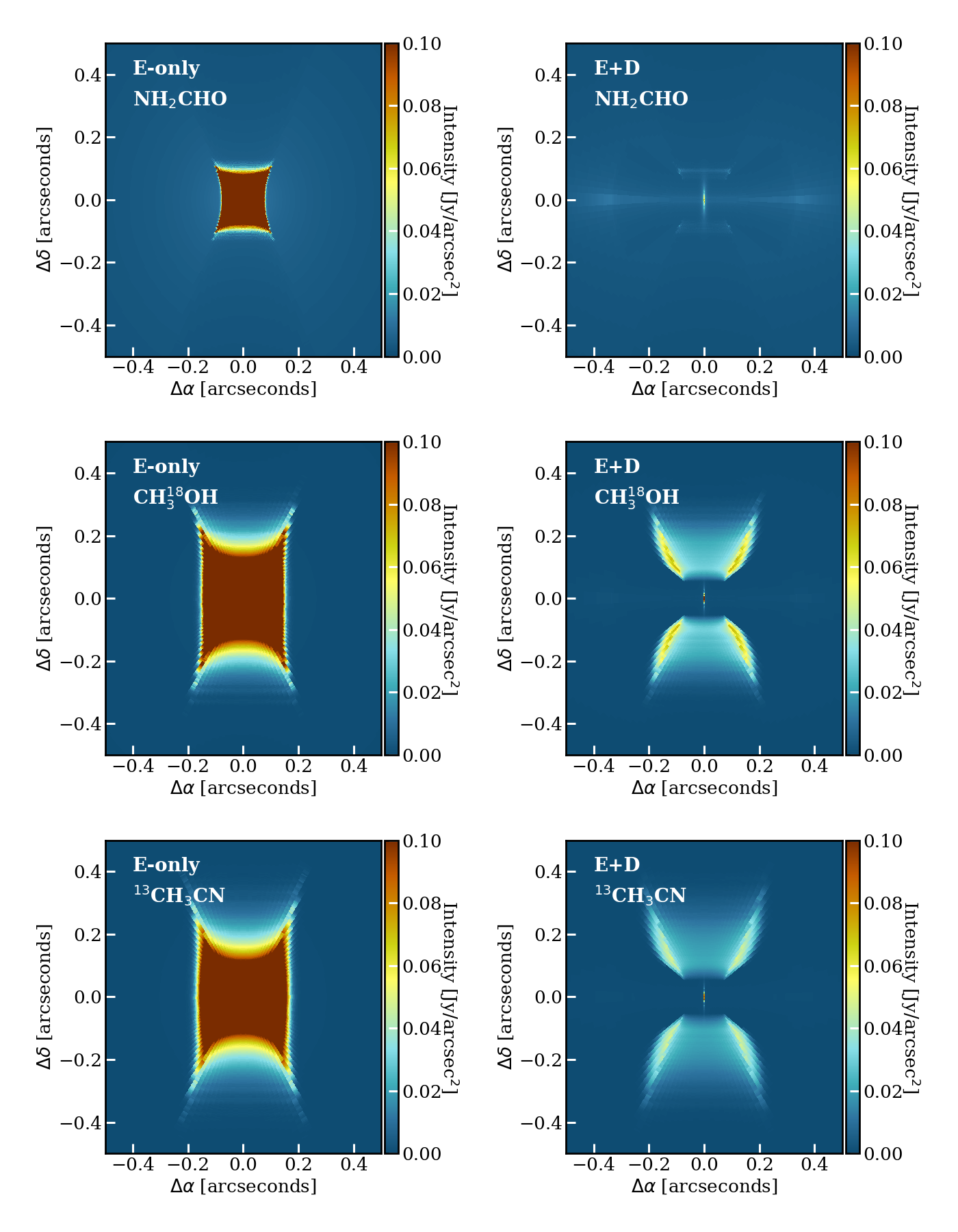}
    \caption{Edge-on images of formamide (top row), methanol (middle), and methyl cyanide (bottom row) at the line emission peak in the low-mass envelope-only (E-only) and envelope-plus-disk (E+D) fiducial models. Dust is not included in these runs to avoid additional emission from dust (in all other runs dust is included unless specified). Here the lines are NH$_2$CHO 14$_{3, 11}-13_{3, 10}$ ($\nu = 299.2552$\,GHz, $E_{\rm up} = 134.1$\,K), CH$_3$OH 10$_{1, 10, 2}-9_{0, 9, 1}$ ($\nu = 326.9612$\,GHz, $E_{\rm up} = 133.1$\,K) run with $^{18}$O abundance, and CH3CN $12_{3, 0} - 11_{-3, 0}$ ($\nu=220.7089$\,GHz and $E_{\rm up} = 133.2$\,K) run with $^{13}$C abundance. Emission from formamide is seen to be less extended than that from methanol and methyl cyanide.} 
    \label{fig:emission}
\end{figure*}

\section{Results}
\label{sec:results}

\subsection{Emitting areas}
\label{sec:emission}

\begin{figure}
  \resizebox{\hsize}{!}{\includegraphics{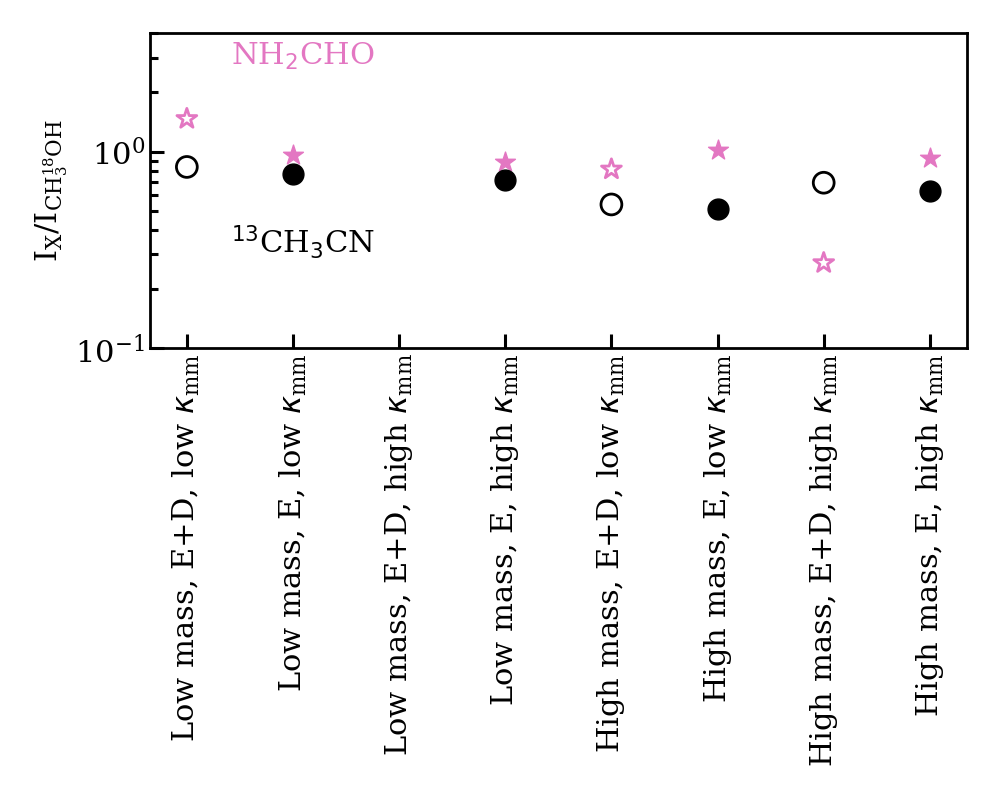}}
  \caption{Integrated line ratios of formamide (pink stars) and methyl cyanide (black circles) to methanol for the fiducial models and those with high mm opacity dust. The lines of formamide, methanol, and methyl cyanide are the same as those in Fig. \ref{fig:emission}. The empty symbols show the models with disks. The considered lines in the low-mass model with disk and high mm opacity dust are not detected and hence there are no measurements. The scatter in NH$_2$CHO/CH$_3^{18}$OH (pink stars) is larger than that in $^{13}$CH$_3$CN/CH$_3^{18}$OH (black circles).}
  \label{fig:line_ratios}
\end{figure}

Figure \ref{fig:emission} presents the modeled emission of NH$_2$CHO, CH$_3^{18}$OH, and $^{13}$CH$_3$CN lines at frequencies of 299.2552\,GHz, 326.9612\,GHz and 220.7089\,GHz which have similar $E_{\rm up}$ (${\sim} 130$\,K) for the fiducial low-mass envelope-only (E-only) and envelope-plus-disk (E+D) models. All emission lines show similar patterns among each other in the envelope-only and envelope-plus-disk models. This is expected because in these models no chemistry is included, while only the balance between adsorption and thermal desorption is considered. In both columns of Fig. \ref{fig:emission} the lines trace the gas around the outflow cavity walls where dust grains are efficiently heated. In the envelope-only models the gas in the envelope on-source also shows emission while this is not the case for the envelope-plus-disk models. Moreover, the emission is fainter in the envelope-plus-disk models, which is particularly obvious for NH$_2$CHO. This is because disk shadowing decreases the temperatures (\citealt{Nazari2022}) and thus less of the molecules are in the gas phase.

As expected formamide has a smaller emitting area than $^{18}$O methanol and $^{13}$C methyl cyanide in envelope-only and envelope-plus-disk models. This is because formamide has a higher binding energy and hence traces the regions closer to the protostar. As Fig. \ref{fig:emission} shows, the difference in sublimation temperature of formamide and methanol results in differences in their emitting areas. That introduces a factor (ratio of the emitting areas) that needs to be considered to convert column density ratios to abundance ratios of these two species. However, as explained in Sect. 4.4.2 of \cite{Nazari2022ALMAGAL} this factor alone does not produce the observed scatter in column density ratios, because if there are no disks around protostars, this factor will be the same between various sources (i.e., the ratio of sublimation temperatures of formamide and methanol is approximately constant between various sources). This factor (emitting area ratios) can only produce a scatter in column density ratios if some sources have a disk and some do not but it is not clear how large this scatter would be. 

\subsection{Effect of disk on line and column density ratios}
\label{sec:column_ratios}

To demonstrate that the difference in emitting areas (Sect. \ref{sec:emission}) can produce a scatter, we first consider examples of line ratios. Figure \ref{fig:line_ratios} presents the integrated line ratios of NH$_2$CHO/CH$_3^{18}$OH and $^{13}$CH$_3$CN/CH$_3^{18}$OH for the fiducial models and those with high mm opacity dust grains. This figure shows that the line ratios of NH$_2$CHO/CH$_3^{18}$OH have a range that covers a factor of ${\sim}6$, while those of $^{13}$CH$_3$CN/CH$_3^{18}$OH cover a range of a factor of ${\sim}1.6$. Moreover, this plot shows that the sources with disks and optically thick dust at mm wavelengths are those that produce the largest scatter for NH$_2$CHO/CH$_3^{18}$OH.

\begin{figure}
    \centering
    \includegraphics[width=\columnwidth]{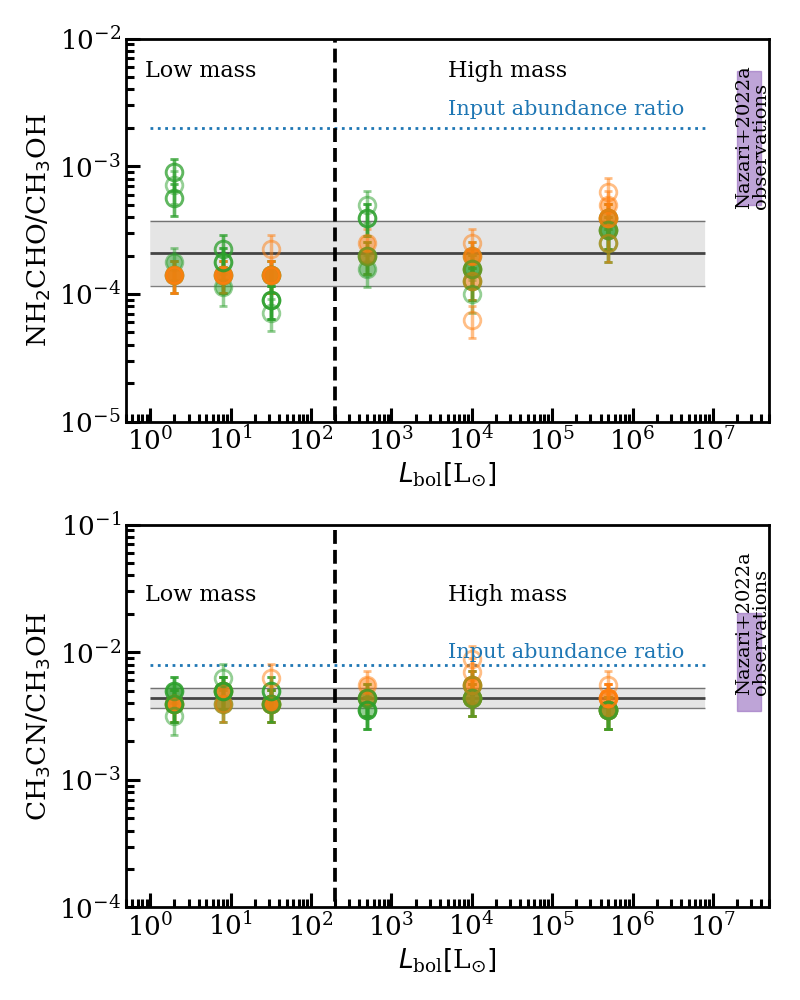}
    \caption{Column density ratios of formamide (top) and methyl cyanide (bottom) to methanol as a function of luminosity for all the models. The solid line shows the weighted mean (by errors) of the $\log_{10}$ of the model points. The gray area presents the weighted standard deviation of the model results. Green shows models with low-mm opacity dust and orange shows those with high-mm opacity dust. Empty circles show the models with a disk and the filled ones present the models without a disk. The purple bar shows one standard deviation below and one standard deviation above the mean of the observational data for low- and high-mass protostars in \cite{Nazari2022ALMAGAL}. The horizontal dotted lines show the abundances assumed in the models. The models with non-detection of methanol are not plotted.} 
    \label{fig:combined}
\end{figure}
We now consider the column densities that are calculated using CASSIS from the radiative transfer models. Appendix \ref{sec:CASSIS_true} demonstrates that as long as dust opacity effects are minimal, the calculated column densities from CASSIS give the correct values of the true total number of molecules assumed in the models within a factor of ${\sim}2$. Moreover, we calculate the column density ratios for the fiducial models with different viewing angles (see Fig. \ref{fig:angles}). At viewing angles larger than zero the line profiles of the disk-plus-envelope models change from singly peaked to triply peaked (Fig. \ref{fig:lines_30_methanol}), with one peak at the transition frequency and two around that which get further from the transition frequency as the inclination angle increases. This occurs due to Keplerian rotation in the disk. For those models our single component CASSIS models fail to reproduce the entire line profile. However, the total flux is dominated by the central peak related to the envelope emission. Therefore, using a multi-component CASSIS fit will not affect our conclusions. We find that the column density ratios for varying viewing angles only change by a factor of $< 2$ (Fig. \ref{fig:angles}). Therefore, for the rest of this paper we only consider the models with a face-on viewing.

Top panel of Fig. \ref{fig:combined} presents the $N_{\rm NH_2CHO}/N_{\rm CH_3OH}$ inferred from the models in this work. The average value of column density ratios of formamide to methanol in Fig. \ref{fig:combined} (top panel) is around one order of magnitude lower than the true abundance ratio assumed in the models which is ${\sim}2 \times 10^{-3}$ (see the horizontal dotted line in Fig. \ref{fig:combined}). This is because the emitting areas of these two species are different (see Fig. \ref{fig:emission}) which results in a correction factor that needs to be multiplied by the column density ratios to give the true abundance. This correction factor is on average ${\sim}10$ for $N_{\rm NH_2CHO}/N_{\rm CH_3OH}$ (Fig. \ref{fig:combined}; also see Sect. \ref{sec:mod_warm_hot} and \ref{sec:correct_obs}). This idea was also suggested in \cite{Nazari2021} for a spherical toy model. For their assumed power-law density and temperature structure they find a correction factor of $(T_{\rm sub, 1}/T_{\rm sub, 2})^{3.75}$, where $T_{\rm sub}$ is the sublimation temperature and subscript 1 and 2 refer to the two molecules in the ratio. Substituting formamide and methanol sublimation temperatures in our models (${\sim}150$\,K and ${\sim}100$\,K) in the above formula gives a correction factor of ${\sim} 4-5$. The larger factor 10 difference between the mean of all the models and the true abundance observed in top panel of Fig. \ref{fig:combined} is likely driven by a combination of effects; inclusion of non-spherical models (i.e., models with disk), inclusion of large dust grains, and the assumption of a fixed temperature when retreiving the column densities (also see Fig. \ref{fig:true_CASSIS}).

\begin{SCfigure*}
    \centering
    \includegraphics[width=11cm]{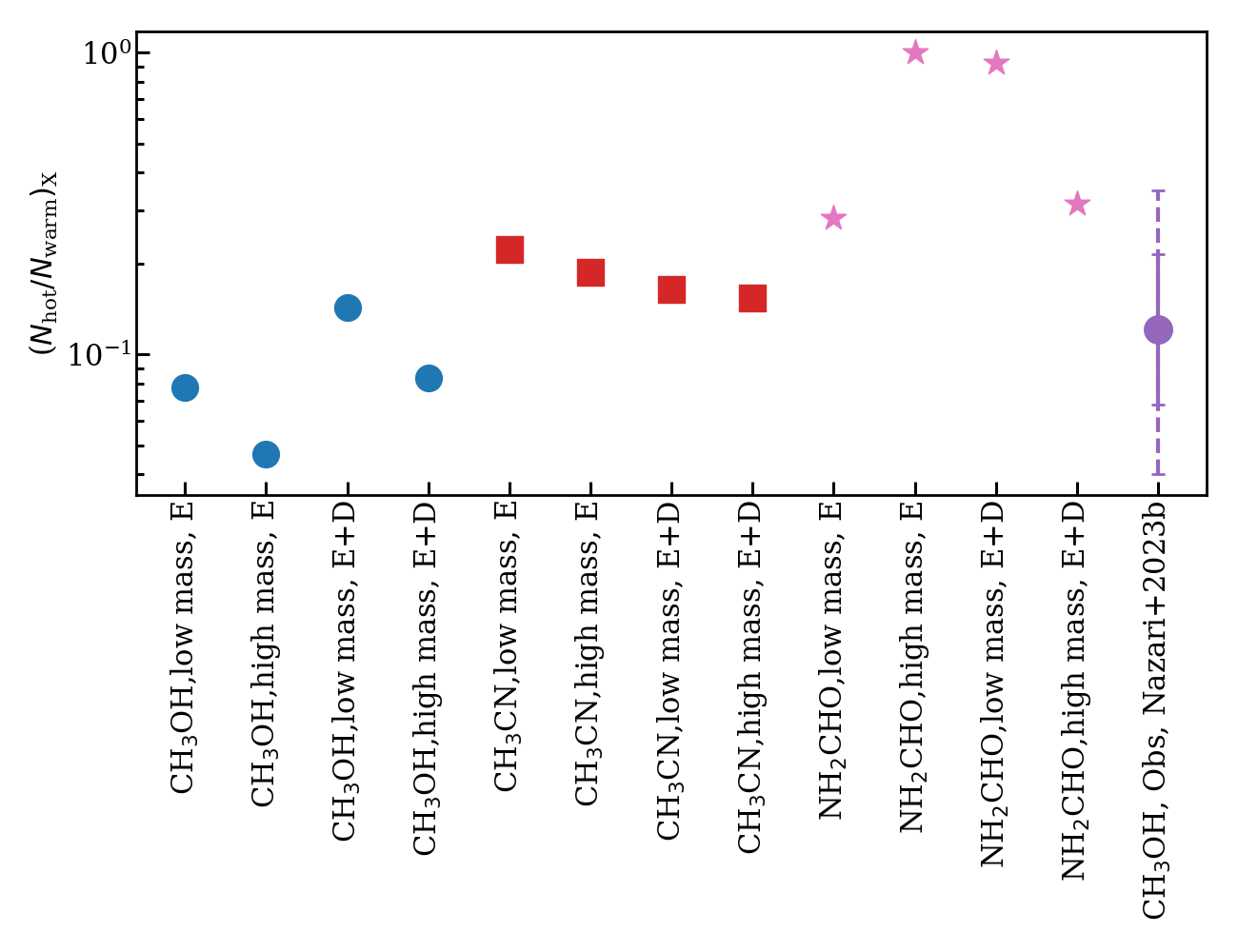}
    \caption{Column density ratios of 100-150\,K gas to those of 300\,K gas for our fiducial envelope-only and envelope-plus-disk models and from observations of \cite{Nazari2023CGD}. The solid error bar shows the standard deviation around the mean of observations while the dashed error bar shows the range in the observational data.} 
    \label{fig:model_hot_warm}
\end{SCfigure*}

Considering the scatter in top panel of Fig. \ref{fig:combined}, it is clear that the scatter from the models is similar to that of observations (purple bar). In other words, models with different physical parameters result in different correction factors with a range of around one order of magnitude. The spread is a factor of 1.8 around the mean from the models in this work while it is a factor of 3.2 from observations of \cite{Nazari2022ALMAGAL} for low- and high-mass protostars. This shows that most of the spread in observations can be explained by the difference in sublimation temperatures of formamide and methanol and the observations only have a factor of 1.78 larger scatter than the models.

It is interesting to note that the models responsible for most of the scatter are those with a disk (empty signs in Fig. \ref{fig:combined}). We also considered how much of the scatter in Fig. \ref{fig:combined} is produced by the dust alone. Figure \ref{fig:dust} presents the ratios of $(\frac{\rm{NH}_2\rm{CHO}}{\rm{CH}_3\rm{OH}})_{\rm{low}\,\kappa_{\rm mm}}/(\frac{\rm{NH}_2\rm{CHO}}{\rm{CH}_3\rm{OH}})_{\rm{high}\,\kappa_{\rm mm}}$, where the numerator and denominator are calculated for the same models with the only difference being dust optical depth. This figure shows that the dust alone can at most produce a factor of 3 scatter but not one order of magnitude. Thus most of the scatter in Fig. \ref{fig:combined} is driven by the variations in source structure. Moreover, it is expected that the scatter is smaller in this work than observations because the models use a constant abundance and the scatter produced here is only affected by physical factors and not chemical considerations. Moreover, if a larger range of disk sizes or envelope masses is used the scatter can be increased. However, the range considered here is representative of the observational results for low- (\citealt{Jorgensen2009}; \citealt{Kristensen2012}; \citealt{Murillo2013}; \citealt{Maury2019}; \citealt{Tobin2020}) and high-mass protostars (\citealt{vanderTak2000}; \citealt{Hunter2014}; \citealt{Johnston2015}; \citealt{Gieser2021}; \citealt{Williams2022}).

We also consider CH$_3$CN as a control molecule due to its similar binding energy to CH$_3$OH. This implies that the correction factor to convert $N_{\rm CH_3CN}/N_{\rm CH_3OH}$ to the true abundance ratio and the scatter in $N_{\rm CH_3CN}/N_{\rm CH_3OH}$ are expected to be minimal. The bottom panel of Fig. \ref{fig:combined} shows that the mean of $N_{\rm CH_3CN}/N_{\rm CH_3OH}$ is only a factor of ${\sim}2$ lower than the assumed abundance in the models (horizontal dotted line). In other words, the correction factor for $N_{\rm CH_3CN}/N_{\rm CH_3OH}$ is ${\sim}2$ compared with ${\sim} 10$ for $N_{\rm NH_2CHO}/N_{\rm CH_3OH}$. This factor of ${\sim}2$ difference is introduced when retrieving the column densities with CASSIS (see Fig. \ref{fig:true_CASSIS}). Moreover, $N_{\rm CH_3CN}/N_{\rm CH_3OH}$ only has a factor of 1.2 scatter around the mean which is a factor of 1.5 smaller than the scatter in $N_{\rm NH_2CHO}/N_{\rm CH_3OH}$ from models. The scatter produced by the models in ratios of $N_{\rm CH_3CN}/N_{\rm CH_3OH}$ is much smaller than that of the observations. More precisely, the spread around the mean is a factor of 1.2 from the models while it is a factor of 2.2 from observations of \cite{Nazari2022ALMAGAL}. That is, the observations have a factor of 1.83 larger scatter than the models. Because of their similar emitting areas, the spread in observations should mainly originate from other effects such as differences in initial ice abundances. Another way to increase the scatter in observations could be potentially different ice environments of CH$_3$CN and CH$_3$OH which will affect the binding energies. In the absence of knowledge on the ice environment of these molecules, the next best approach is the use of high-angular resolution data to accurately measure the difference between sublimation regions of these molecules.

Finally, we considered the effect of our assumed binding energies on the conclusions. We varied the binding energies in our fiducial models by taking three values for each molecule representing the range of binding energies reported in the literature (denoted as low, medium, and high). Fig. \ref{fig:min_max} presents the column density ratios for these three combinations. This figure shows that the difference between the column density ratios of CH$_3$CN/CH$_3$OH for the range of binding energies reported in the literature is within a factor of around 3. This is expected because in all the considered cases methanol and methyl cyanide have similar binding energies. However, the ratio of NH$_2$CHO/CH$_3$OH can change by one order of magnitude depending on which binding energy is assumed. This shows that depending on the ice matrix that formamide and methanol reside in (which affects their binding energies), the scatter in column density ratios could be easily affected. 

\subsection{Temperature components in models}
\label{sec:mod_warm_hot}

Here we consider a two-component temperature fit to find the warm (${\sim}100$\,K) and hot (${\sim}300$\,K) column densities in the fiducial models by fitting the low- and high-$E_{\rm up}$ lines separately. We first fit all the lines with a range of $E_{\rm up}$ (see the lines with and without stars in Table \ref{tab:lines}) with temperatures of 100-150\,K (warm gas). Then on top of that we add a hot component with temperature of 300\,K to complete the fit for the lines with $E_{\rm up}>500$\,K . We then fine tune the fits to make sure that the sum of the two components gives a good fit to all the emission lines. This is to consider the column densities of target species in the hotter regions closer to the protostar.

Although variations between column densities for the assumed excitation temperatures that differ by ${\sim}50$\,K is not significant (a factor of $\lesssim 2$), the second hot component fitted at 300\,K can be significantly different from the warm component fitted at 100-150\,K. Figure \ref{fig:model_hot_warm} presents the ratio of column densities for the hot gas (300\,K) to those of the warm gas (100-150\,K). In particular, Fig. \ref{fig:model_hot_warm} shows that methanol and methyl cyanide have ${\sim}1$ order of magnitude lower hot component than warm component. This suggests that if a single temperature were used to fit the lines and find the column densities of these two molecules in spatially unresolved observations, the temperature and the column density associated with it would likely be dominated by the low-$E_{\rm up}$ lines. This results in a column density that traces the warm regions further from the protostar. 

We note that this drop in hot column densities should only occur if a constant abundance is assumed for these species in the gas with no further gas-phase chemistry. \cite{Nazari2023CGD} found that methanol, on average, has ${\sim}1$ order of magnitude lower hot column densities agreeing well with our models (see the purple data point in Fig. \ref{fig:model_hot_warm}). However, they found that this drop in hot column densities does not happen for methyl cyanide where additional hot gas-phase chemistry is needed as a result of destruction of refractory organics to explain the observations.

The behavior of formamide is different from the other two molecules because it has a higher sublimation temperature (by ${\sim}40-50$\,K in our models). The sublimation temperatures of methanol and methyl cyanide in our models are around 100\,K while this value for formamide is around 150\,K. Therefore, at low temperatures formamide has not completely desorbed from the ices. Figure \ref{fig:model_hot_warm} shows that formamide has at most a factor of ${\sim}3$ drop in its hot column densities compared with its warm column densities, while this drop was a factor of ${\gtrsim}10$ for methanol. This is because of formamide's higher sublimation temperature and hence pointing to the fact that column densities found from spatially unresolved observations of this molecule with a single-temperature fit will be equally dominated by its hot and warm components.

To conclude, if no gas-phase chemistry occurs (assumption of constant abundances in our models) molecules with similar sublimation temperatures to methanol are expected to have around 1 order of magnitude lower hot (${\sim}300$\,K) column densities than warm (${\sim}100$\,K) column densities. In other words, the warm column densities are expected to dominate the total column density for methanol, while the warm and hot column densities of formamide contribute roughly equally to the total column density.

\section{Discussion}
\label{sec:discussion}

\subsection{Correcting for difference in emitting areas in observations}
\label{sec:correct_obs}

Here we consider the methanol emission that is coming from a similar region to formamide. Based on Sect. \ref{sec:mod_warm_hot}, observational results for formamide and methanol, which are found by either fixing the excitation temperature or fitting the lines with a single temperature, should be biased to either the hot or the warm component of each molecule. In the models, formamide is a molecule with similar column densities for its hot and warm components while methanol is a molecule whose column density is dominated by its warm component if a single-component analysis is used. The bias produced from the different binding energies in the scatter of $N_{\rm NH_2CHO}/N_{\rm CH_3OH}$ could be corrected if the ratio of the warm column density of formamide (which is similar to its hot component) to the hot column density of methanol are found. This is because the hot methanol is expected to trace the regions closer to the protostar, similar to the regions that formamide traces.

\begin{SCfigure*}
    \centering
    \includegraphics[width=11cm]{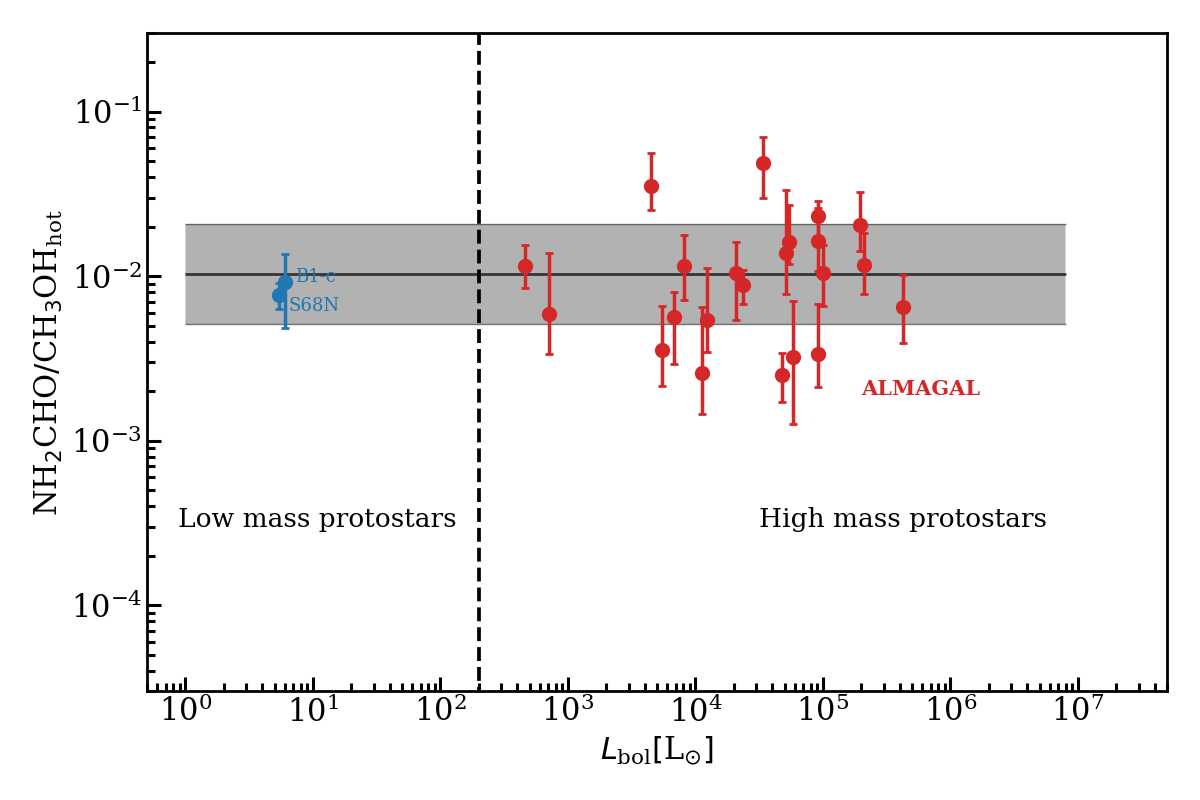}
    \caption{Column density ratio of formamide to hot methanol. The red signs show the ALMAGAL sources and the blue signs are the two low-mass sources B1-c and S68N. The gray area shows the weighted standard deviation of $log_{10}$ of the data points around the weighted mean of $log_{10}$ of the data points (black solid line). The hot column densities for methanol are taken from \cite{Nazari2023CGD} and the column densities of formamide are taken from \citet{Nazari2021,Nazari2022ALMAGAL}. When focusing on methanol only at high temperatures closer to the sublimation temperature of formamide, the scatter is reduced.} 
    \label{fig:hot_ratio}
\end{SCfigure*}

Figure \ref{fig:hot_ratio} presents formamide to hot methanol for low- and high-mass protostars. The hot methanol column densities for sources from the ALMA Evolutionary study of High Mass Protocluster Formation in the Galaxy (ALMAGAL) are taken from \cite{Nazari2023CGD} and those for B1-c and S68N are calculated in this work using the same spectra of \cite{vanGelder2020}. For these two sources the column densities are found from fitting the high-$E_{\rm up}$ lines of major methanol isotopologue using a similar method to \cite{Nazari2023CGD}. 

In Fig. \ref{fig:hot_ratio} the ratio of formamide to hot methanol has a mean (${\sim} 10^{-2}$) that is about one order of magnitude higher than that in \cite{Nazari2022ALMAGAL} found for the ratio of formamide to methanol in a single-temperature component analysis (${\sim}2 \times 10^{-3}$). This is expected from our models where the hot component of methanol is ${\sim}1$ order of magnitude lower than its warm component (Sect. \ref{sec:mod_warm_hot}). Moreover, this higher value is more consistent with what is found in comets for ratio of NH$_2$CHO/CH$_3$OH (\citealt{LeRoy2015}; \citealt{Altwegg2019}). This emphasizes once more that the gas column density ratios derived with a single-temperature analysis might not represent the true ice abundances for molecules with different sublimation temperatures. This is also shown in Fig. \ref{fig:combined} where the true abundance assumed in the models for NH$_2$CHO/CH$_3$OH is ${\sim}1$ order of magnitude higher than that measured for gas-phase column density ratios in the models.

The range of data points in Fig. \ref{fig:hot_ratio} is around one order of magnitude while this range in ALMAGAL sources was around two orders of magnitude from single-temperature fitting. Furthermore, the observational scatter found in Fig. \ref{fig:hot_ratio} is a factor of 2.0 around the mean. This is a factor of 1.6 smaller than what was previously found for the formamide to methanol ratio in the ALMAGAL sample. The new scatter found for the ratio of formamide to methanol (factor 2.0) is similar to the scatter found for other COM ratios in \cite{Nazari2022ALMAGAL}, so that formamide no longer stands out as having an exceptionally high scatter, making it a `normal' molecule.

\subsection{Implications for formamide formation}

Figure \ref{fig:combined} shows and Sect. \ref{sec:column_ratios} explains that the bulk of the observed spread in $N_{\rm NH_2CHO}/N_{\rm CH_3OH}$ can be described by our models with a range of physical structures. Chemical networks are not included in this work and all the abundances are parametrized. Therefore, the spread in column density ratios is only produced by the difference between methanol and formamide sublimation temperatures. Our conclusions agree with the observational findings of \cite{Suzuki2018} who find a stronger correlation among N-bearing COMs than among N- and O-bearing species toward eight star forming regions. This is interesting given that N-bearing COMs, although not always, but on average have higher binding energies than O-bearing ones.

Finally, Sect. \ref{sec:correct_obs} shows that if the difference in sublimation temperatures of formamide and methanol is accounted for, the scatter in the observations decreases by a factor of 1.6. Hence the scatters in column density ratios as large as those seen in ratios of formamide with respect to other molecules (\citealt{Nazari2022ALMAGAL}) do not necessarily imply that gas-phase formation routes are effective. Although the effect of gas-phase formation routes on column density ratios cannot be excluded, the bulk of scatter can be simply explained by physical effects such as existence of disks with varying sizes around protostars. Therefore, formamide could be forming in the prestellar ices along with other complex organics.

\section{Conclusions}
\label{sec:conclusions}

In this work we investigate the influence of physical factors on observables such as column density ratios. We modeled the emission of formamide, methanol and methyl cyanide by parametrizing their abundances using RADMC3D in models with and without a disk. Our models encompass a large range of physical parameters. The column densities of these species are calculated for each model in the same way as done in real observations. Our main conclusions are summarized below:

\begin{itemize}
    \item The emitting area of formamide is smaller than that of methanol and methyl cyanide in our models because of their distinct binding energies (and sublimation temperatures).
    \item The column density ratios of formamide to methanol from the models, especially those with varying disk sizes, produce a scatter comparable with those of observations (Fig. \ref{fig:combined}). This suggests that a large part of the scatter seen in column density ratios of formamide to methanol could be only due to physical effects.  
    \item The scatter in column density ratios of methyl cyanide to methanol from the models is much smaller than that of formamide to methanol. It is also ${\sim} 2$ times lower than that of observations. This confirms that the observational scatter in column density ratios of two molecules with similar binding energies most probably has a chemical origin or points to those molecules residing in different ice matrices.
    \item We find that varying the binding energies within the ranges suggested in the literature can change the ratios of CH$_3$CN/CH$_3$OH by a factor of about 3 while that can change the ratios of NH$_2$CHO/CH$_3$OH by up to a factor of 10. This emphasizes the importance of having information on the ice environment and thus the binding energies of these molecules for more robust conclusions on chemical formation pathways.   
    \item A two-component temperature analysis reveals that formamide has similar hot (300\,K) and warm (100-150\,K) column densities while methanol is dominated by its warm component. We find ${\sim}1$ order of magnitude lower hot than warm methanol in the models with constant gas-phase abundances, consistent with the observations.
    \item We correct for the difference in sublimation temperatures of formamide and methanol in observations (Fig. \ref{fig:hot_ratio}). After correction the scatter in observations decreases by a factor of 1.6, making formamide a molecule similar to the other COMs. Therefore, formamide could also be formed in the prestellar phase on grains along with the other species. However, this conclusion does not exclude the potential effects of chemistry (in the gas or on grains) on the produced scatter.
    \item The corrected column density ratio of $N_{\rm NH_2CHO}/N_{\rm CH_3OH, hot}$ has a mean comparable to those of comets. This highlights that gas-phase column density ratios measured assuming a single temperature for species with sublimation temperatures as different as methanol and formamide could be off from the true ice abundance ratios by ${\sim}1$ order of magnitude (also see Fig. \ref{fig:combined}). 
\end{itemize} 

This work shows that if two molecules have different sublimation temperatures (i.e., different emitting sizes) the mean and scatter in column density ratios inferred from observations will be affected by that. This has great implications for ice observations of complex organics by the \textit{James Webb Space Telescope} (JWST). For example, formamide is a molecule with gas-phase column density ratios with respect to methanol that are below the sensitivity limit of JWST while the true abundance ratio of NH$_2$CHO/CH$_3$OH in ices could be ${\sim}1$ order of magnitude higher, making it more probable to be detected by JWST (see \citealt{Slavicinska2023} for JWST results). Finally, an example of different emitting sizes of formamide and methanol is already observed toward the low-mass protostar HH 212 (\citealt{Lee2022}). The next step is to directly measure such differences with high-angular resolution ALMA data in a larger sample of sources.

\begin{acknowledgements}
    We thank the referee for the constructive comments. We also thank N. F. W. Ligterink for the helpful discussions. Astrochemistry in Leiden is supported by the Netherlands Research School for Astronomy (NOVA), by funding from the European Research Council (ERC) under the European Union’s Horizon 2020 research and innovation programme (grant agreement No. 101019751 MOLDISK), and by the Dutch Research Council (NWO) grant 618.000.001. Support by the Danish National Research Foundation through the Center of Excellence “InterCat” (Grant agreement no.: DNRF150) is also acknowledged. B.T. acknowledges support from the Programme National ``Physique et Chimie du Milieu Interstellair'' (PCMI) of CNRS/INSU with INC/INP and co-funded by CNES. G.P.R. acknowledges support from the Netherlands Organisation for Scientific Research (NWO, program number 016.Veni.192.233) and from an STFC Ernest Rutherford Fellowship (grant number ST/T003855/1). This project has received funding from the European Research Council (ERC) under the European Union's Horizon Europe Research \& Innovation Programme under grant agreement No 101039651 (DiscEvol).   
\end{acknowledgements}

%
%

\bibliographystyle{aa}
\bibliography{paper_ratios}

\begin{thebibliography}{81}
\expandafter\ifx\csname natexlab\endcsname\relax\def\natexlab#1{#1}\fi

\bibitem[{{Allen} {et~al.}(2020){Allen}, {van der Tak}, {L{\'o}pez-Sepulcre}, {S{\'a}nchez-Monge}, {Rivilla}, \& {Cesaroni}}]{Allen2020}
{Allen}, V., {van der Tak}, F.~F.~S., {L{\'o}pez-Sepulcre}, A., {et~al.} 2020, \aap, 636, A67

\bibitem[{{Altwegg} {et~al.}(2019){Altwegg}, {Balsiger}, \& {Fuselier}}]{Altwegg2019}
{Altwegg}, K., {Balsiger}, H., \& {Fuselier}, S.~A. 2019, \araa, 57, 113

\bibitem[{{Baek} {et~al.}(2022){Baek}, {Lee}, {Hirota}, {Kim}, \& {Kyoung Kim}}]{Baek2022}
{Baek}, G., {Lee}, J.-E., {Hirota}, T., {Kim}, K.-T., \& {Kyoung Kim}, M. 2022, \apj, 939, 84

\bibitem[{{Barone} {et~al.}(2015){Barone}, {Latouche}, {Skouteris}, {Vazart}, {Balucani}, {Ceccarelli}, \& {Lefloch}}]{Barone2015}
{Barone}, V., {Latouche}, C., {Skouteris}, D., {et~al.} 2015, \mnras, 453, L31

\bibitem[{{Belloche} {et~al.}(2020){Belloche}, {Maury}, {Maret}, {Anderl}, {Bacmann}, {Andr{\'e}}, {Bontemps}, {Cabrit}, {Codella}, {Gaudel}, {Gueth}, {Lef{\`e}vre}, {Lefloch}, {Podio}, \& {Testi}}]{Belloche2020}
{Belloche}, A., {Maury}, A.~J., {Maret}, S., {et~al.} 2020, \aap, 635, A198

\bibitem[{{Beltr{\'a}n} {et~al.}(2009){Beltr{\'a}n}, {Codella}, {Viti}, {Neri}, \& {Cesaroni}}]{Beltran2009}
{Beltr{\'a}n}, M.~T., {Codella}, C., {Viti}, S., {Neri}, R., \& {Cesaroni}, R. 2009, \apjl, 690, L93

\bibitem[{{Beuther} {et~al.}(2017){Beuther}, {Walsh}, {Johnston}, {Henning}, {Kuiper}, {Longmore}, \& {Walmsley}}]{Beuther2017}
{Beuther}, H., {Walsh}, A.~J., {Johnston}, K.~G., {et~al.} 2017, \aap, 603, A10

\bibitem[{{Blake} {et~al.}(1987){Blake}, {Sutton}, {Masson}, \& {Phillips}}]{Blake1987}
{Blake}, G.~A., {Sutton}, E.~C., {Masson}, C.~R., \& {Phillips}, T.~G. 1987, \apj, 315, 621

\bibitem[{{Boogert} {et~al.}(2015){Boogert}, {Gerakines}, \& {Whittet}}]{Boogert2015}
{Boogert}, A.~C.~A., {Gerakines}, P.~A., \& {Whittet}, D. C.~B. 2015, \araa, 53, 541

\bibitem[{{Busch} {et~al.}(2022){Busch}, {Belloche}, {Garrod}, {M{\"u}ller}, \& {Menten}}]{Busch2022}
{Busch}, L.~A., {Belloche}, A., {Garrod}, R.~T., {M{\"u}ller}, H. S.~P., \& {Menten}, K.~M. 2022, \aap, 665, A96

\bibitem[{{Cazzoli} \& {Puzzarini}(2006)}]{Cazzoli2006}
{Cazzoli}, G. \& {Puzzarini}, C. 2006, Journal of Molecular Spectroscopy, 240, 153

\bibitem[{{Ceccarelli} {et~al.}(2017){Ceccarelli}, {Caselli}, {Fontani}, {Neri}, {L{\'o}pez-Sepulcre}, {Codella}, {Feng}, {Jim{\'e}nez-Serra}, {Lefloch}, {Pineda}, {Vastel}, {Alves}, {Bachiller}, {Balucani}, {Bianchi}, {Bizzocchi}, {Bottinelli}, {Caux}, {Chac{\'o}n-Tanarro}, {Choudhury}, {Coutens}, {Dulieu}, {Favre}, {Hily-Blant}, {Holdship}, {Kahane}, {Jaber Al-Edhari}, {Laas}, {Ospina}, {Oya}, {Podio}, {Pon}, {Punanova}, {Quenard}, {Rimola}, {Sakai}, {Sims}, {Spezzano}, {Taquet}, {Testi}, {Theul{\'e}}, {Ugliengo}, {Vasyunin}, {Viti}, {Wiesenfeld}, \& {Yamamoto}}]{Ceccarelli2017}
{Ceccarelli}, C., {Caselli}, P., {Fontani}, F., {et~al.} 2017, \apj, 850, 176

\bibitem[{{Chaabouni} {et~al.}(2018){Chaabouni}, {Diana}, {Nguyen}, \& {Dulieu}}]{Chaabouni2018}
{Chaabouni}, H., {Diana}, S., {Nguyen}, T., \& {Dulieu}, F. 2018, \aap, 612, A47

\bibitem[{{Chahine} {et~al.}(2022){Chahine}, {L{\'o}pez-Sepulcre}, {Neri}, {Ceccarelli}, {Mercimek}, {Codella}, {Bouvier}, {Bianchi}, {Favre}, {Podio}, {Alves}, {Sakai}, \& {Yamamoto}}]{Chahine2022}
{Chahine}, L., {L{\'o}pez-Sepulcre}, A., {Neri}, R., {et~al.} 2022, \aap, 657, A78

\bibitem[{{Chen} {et~al.}(2023){Chen}, {van Gelder}, {Nazari}, {McGuire}, {Brogan}, \& et~al.}]{Chen2023}
{Chen}, Y., {van Gelder}, M.~L., {Nazari}, P., {et~al.} 2023, Submitted to A\&A

\bibitem[{{Codella} {et~al.}(2017){Codella}, {Ceccarelli}, {Caselli}, {Balucani}, {Barone}, {Fontani}, {Lefloch}, {Podio}, {Viti}, {Feng}, {Bachiller}, {Bianchi}, {Dulieu}, {Jim{\'e}nez-Serra}, {Holdship}, {Neri}, {Pineda}, {Pon}, {Sims}, {Spezzano}, {Vasyunin}, {Alves}, {Bizzocchi}, {Bottinelli}, {Caux}, {Chac{\'o}n-Tanarro}, {Choudhury}, {Coutens}, {Favre}, {Hily-Blant}, {Kahane}, {Jaber Al-Edhari}, {Laas}, {L{\'o}pez-Sepulcre}, {Ospina}, {Oya}, {Punanova}, {Puzzarini}, {Quenard}, {Rimola}, {Sakai}, {Skouteris}, {Taquet}, {Testi}, {Theul{\'e}}, {Ugliengo}, {Vastel}, {Vazart}, {Wiesenfeld}, \& {Yamamoto}}]{Codella2017}
{Codella}, C., {Ceccarelli}, C., {Caselli}, P., {et~al.} 2017, \aap, 605, L3

\bibitem[{{Codella} {et~al.}(2022){Codella}, {L{\'o}pez-Sepulcre}, {Ohashi}, {Chandler}, {De Simone}, {Podio}, {Ceccarelli}, {Sakai}, {Alves}, {Dur{\'a}n}, {Fedele}, {Loinard}, {Mercimek}, {Murillo}, {Zhang}, {Bianchi}, {Bouvier}, {Busquet}, {Caselli}, {Dulieu}, {Feng}, {Hanawa}, {Johnstone}, {Lefloch}, {Maud}, {Moellenbrock}, {Oya}, {Svoboda}, \& {Yamamoto}}]{Codella2022}
{Codella}, C., {L{\'o}pez-Sepulcre}, A., {Ohashi}, S., {et~al.} 2022, \mnras, 515, 543

\bibitem[{{Coletta} {et~al.}(2020){Coletta}, {Fontani}, {Rivilla}, {Mininni}, {Colzi}, {S{\'a}nchez-Monge}, \& {Beltr{\'a}n}}]{Coletta2020}
{Coletta}, A., {Fontani}, F., {Rivilla}, V.~M., {et~al.} 2020, \aap, 641, A54

\bibitem[{{Coutens} {et~al.}(2016){Coutens}, {J{\o}rgensen}, {van der Wiel}, {M{\"u}ller}, {Lykke}, {Bjerkeli}, {Bourke}, {Calcutt}, {Drozdovskaya}, {Favre}, {Fayolle}, {Garrod}, {Jacobsen}, {Ligterink}, {{\"O}berg}, {Persson}, {van Dishoeck}, \& {Wampfler}}]{Coutens2016}
{Coutens}, A., {J{\o}rgensen}, J.~K., {van der Wiel}, M.~H.~D., {et~al.} 2016, \aap, 590, L6

\bibitem[{{Douglas} {et~al.}(2022){Douglas}, {Lucas}, {Walsh}, {West}, {Blitz}, \& {Heard}}]{Douglas2022}
{Douglas}, K.~M., {Lucas}, D.~I., {Walsh}, C., {et~al.} 2022, \apjl, 937, L16

\bibitem[{{Dullemond} {et~al.}(2012){Dullemond}, {Juhasz}, {Pohl}, {Sereshti}, {Shetty}, {Peters}, {Commercon}, \& {Flock}}]{Dullemond2012}
{Dullemond}, C.~P., {Juhasz}, A., {Pohl}, A., {et~al.} 2012, {RADMC-3D: A multi-purpose radiative transfer tool}, Astrophysics Source Code Library, record ascl:1202.015

\bibitem[{{Ferrero} {et~al.}(2020){Ferrero}, {Zamirri}, {Ceccarelli}, {Witzel}, {Rimola}, \& {Ugliengo}}]{Ferrero2020}
{Ferrero}, S., {Zamirri}, L., {Ceccarelli}, C., {et~al.} 2020, \apj, 904, 11

\bibitem[{{Fuchs} {et~al.}(2009){Fuchs}, {Cuppen}, {Ioppolo}, {Romanzin}, {Bisschop}, {Andersson}, {van Dishoeck}, \& {Linnartz}}]{Fuchs2009}
{Fuchs}, G.~W., {Cuppen}, H.~M., {Ioppolo}, S., {et~al.} 2009, \aap, 505, 629

\bibitem[{{Garrod} {et~al.}(2022){Garrod}, {Jin}, {Matis}, {Jones}, {Willis}, \& {Herbst}}]{Garrod2022}
{Garrod}, R.~T., {Jin}, M., {Matis}, K.~A., {et~al.} 2022, \apjs, 259, 1

\bibitem[{{Gieser} {et~al.}(2021){Gieser}, {Beuther}, {Semenov}, {Ahmadi}, {Suri}, {M{\"o}ller}, {Beltr{\'a}n}, {Klaassen}, {Zhang}, {Urquhart}, {Henning}, {Feng}, {Galv{\'a}n-Madrid}, {de Souza Magalh{\~a}es}, {Moscadelli}, {Longmore}, {Leurini}, {Kuiper}, {Peters}, {Menten}, {Csengeri}, {Fuller}, {Wyrowski}, {Lumsden}, {S{\'a}nchez-Monge}, {Maud}, {Linz}, {Palau}, {Schilke}, {Pety}, {Pudritz}, {Winters}, \& {Pi{\'e}tu}}]{Gieser2021}
{Gieser}, C., {Beuther}, H., {Semenov}, D., {et~al.} 2021, \aap, 648, A66

\bibitem[{{Hasegawa} {et~al.}(1992){Hasegawa}, {Herbst}, \& {Leung}}]{Hasegawa1992}
{Hasegawa}, T.~I., {Herbst}, E., \& {Leung}, C.~M. 1992, \apjs, 82, 167

\bibitem[{Haupa {et~al.}(2019)Haupa, Tarczay, \& Lee}]{Haupa2019}
Haupa, K.~A., Tarczay, G., \& Lee, Y.-P. 2019, Journal of the American Chemical Society, 141, 11614, pMID: 31246013

\bibitem[{{Herbst} \& {van Dishoeck}(2009)}]{Herbst2009}
{Herbst}, E. \& {van Dishoeck}, E.~F. 2009, \araa, 47, 427

\bibitem[{{Hosokawa} \& {Omukai}(2009)}]{Hosokawa2009}
{Hosokawa}, T. \& {Omukai}, K. 2009, \apj, 691, 823

\bibitem[{{Hunter} {et~al.}(2014){Hunter}, {Brogan}, {Cyganowski}, \& {Young}}]{Hunter2014}
{Hunter}, T.~R., {Brogan}, C.~L., {Cyganowski}, C.~J., \& {Young}, K.~H. 2014, \apj, 788, 187

\bibitem[{{Johnston} {et~al.}(2015){Johnston}, {Robitaille}, {Beuther}, {Linz}, {Boley}, {Kuiper}, {Keto}, {Hoare}, \& {van Boekel}}]{Johnston2015}
{Johnston}, K.~G., {Robitaille}, T.~P., {Beuther}, H., {et~al.} 2015, \apjl, 813, L19

\bibitem[{{Jones} {et~al.}(2011){Jones}, {Bennett}, \& {Kaiser}}]{Jones2011}
{Jones}, B.~M., {Bennett}, C.~J., \& {Kaiser}, R.~I. 2011, \apj, 734, 78

\bibitem[{{J{\o}rgensen} {et~al.}(2016){J{\o}rgensen}, {van der Wiel}, {Coutens}, {Lykke}, {M{\"u}ller}, {van Dishoeck}, {Calcutt}, {Bjerkeli}, {Bourke}, {Drozdovskaya}, {Favre}, {Fayolle}, {Garrod}, {Jacobsen}, {{\"O}berg}, {Persson}, \& {Wampfler}}]{Jorgensen2016}
{J{\o}rgensen}, J.~K., {van der Wiel}, M.~H.~D., {Coutens}, A., {et~al.} 2016, \aap, 595, A117

\bibitem[{{J{\o}rgensen} {et~al.}(2009){J{\o}rgensen}, {van Dishoeck}, {Visser}, {Bourke}, {Wilner}, {Lommen}, {Hogerheijde}, \& {Myers}}]{Jorgensen2009}
{J{\o}rgensen}, J.~K., {van Dishoeck}, E.~F., {Visser}, R., {et~al.} 2009, \aap, 507, 861

\bibitem[{{Kristensen} {et~al.}(2012){Kristensen}, {van Dishoeck}, {Bergin}, {Visser}, {Y{\i}ld{\i}z}, {San Jose-Garcia}, {J{\o}rgensen}, {Herczeg}, {Johnstone}, {Wampfler}, {Benz}, {Bruderer}, {Cabrit}, {Caselli}, {Doty}, {Harsono}, {Herpin}, {Hogerheijde}, {Karska}, {van Kempen}, {Liseau}, {Nisini}, {Tafalla}, {van der Tak}, \& {Wyrowski}}]{Kristensen2012}
{Kristensen}, L.~E., {van Dishoeck}, E.~F., {Bergin}, E.~A., {et~al.} 2012, \aap, 542, A8

\bibitem[{{Kukolich} \& {Nelson}(1971)}]{Kukolich1971}
{Kukolich}, S.~G. \& {Nelson}, A.~C. 1971, Chemical Physics Letters, 11, 383

\bibitem[{{Law} {et~al.}(2021){Law}, {Zhang}, {{\"O}berg}, {Galv{\'a}n-Madrid}, {Keto}, {Liu}, \& {Ho}}]{Law2021}
{Law}, C.~J., {Zhang}, Q., {{\"O}berg}, K.~I., {et~al.} 2021, \apj, 909, 214

\bibitem[{{Le Roy} {et~al.}(2015){Le Roy}, {Altwegg}, {Balsiger}, {Berthelier}, {Bieler}, {Briois}, {Calmonte}, {Combi}, {De Keyser}, {Dhooghe}, {Fiethe}, {Fuselier}, {Gasc}, {Gombosi}, {H{\"a}ssig}, {J{\"a}ckel}, {Rubin}, \& {Tzou}}]{LeRoy2015}
{Le Roy}, L., {Altwegg}, K., {Balsiger}, H., {et~al.} 2015, \aap, 583, A1

\bibitem[{{Lee} {et~al.}(2022){Lee}, {Codella}, {Ceccarelli}, \& {L{\'o}pez-Sepulcre}}]{Lee2022}
{Lee}, C.-F., {Codella}, C., {Ceccarelli}, C., \& {L{\'o}pez-Sepulcre}, A. 2022, \apj, 937, 10

\bibitem[{{Ligterink} {et~al.}(2021){Ligterink}, {Ahmadi}, {Coutens}, {Tychoniec}, {Calcutt}, {van Dishoeck}, {Linnartz}, {J{\o}rgensen}, {Garrod}, \& {Bouwman}}]{Ligterink2021}
{Ligterink}, N.~F.~W., {Ahmadi}, A., {Coutens}, A., {et~al.} 2021, \aap, 647, A87

\bibitem[{{Ligterink} {et~al.}(2022){Ligterink}, {Ahmadi}, {Luitel}, {Coutens}, {Calcutt}, {Tychoniec}, {Linnartz}, {J{\o}rgensen}, {Garrod}, \& {Bouwman}}]{Ligterink2022}
{Ligterink}, N. F.~W., {Ahmadi}, A., {Luitel}, B., {et~al.} 2022, ACS Earth and Space Chemistry, 6, 455

\bibitem[{{Ligterink} {et~al.}(2020){Ligterink}, {El-Abd}, {Brogan}, {Hunter}, {Remijan}, {Garrod}, \& {McGuire}}]{Ligterink2020}
{Ligterink}, N. F.~W., {El-Abd}, S.~J., {Brogan}, C.~L., {et~al.} 2020, \apj, 901, 37

\bibitem[{{Ligterink} \& {Minissale}(2023)}]{Ligterink2023}
{Ligterink}, N.~F.~W. \& {Minissale}, M. 2023, \aap, 676, A80

\bibitem[{{Mart{\'\i}n-Dom{\'e}nech} {et~al.}(2021){Mart{\'\i}n-Dom{\'e}nech}, {Bergner}, {{\"O}berg}, {Carpenter}, {Law}, {Huang}, {J{\o}rgensen}, {Schwarz}, \& {Wilner}}]{Martin2021}
{Mart{\'\i}n-Dom{\'e}nech}, R., {Bergner}, J.~B., {{\"O}berg}, K.~I., {et~al.} 2021, \apj, 923, 155

\bibitem[{{Maury} {et~al.}(2019){Maury}, {Andr{\'e}}, {Testi}, {Maret}, {Belloche}, {Hennebelle}, {Cabrit}, {Codella}, {Gueth}, {Podio}, {Anderl}, {Bacmann}, {Bontemps}, {Gaudel}, {Ladjelate}, {Lef{\`e}vre}, {Tabone}, \& {Lefloch}}]{Maury2019}
{Maury}, A.~J., {Andr{\'e}}, P., {Testi}, L., {et~al.} 2019, \aap, 621, A76

\bibitem[{{McGuire}(2022)}]{McGuire2021}
{McGuire}, B.~A. 2022, \apjs, 259, 30

\bibitem[{{Milam} {et~al.}(2005){Milam}, {Savage}, {Brewster}, {Ziurys}, \& {Wyckoff}}]{Milam2005}
{Milam}, S.~N., {Savage}, C., {Brewster}, M.~A., {Ziurys}, L.~M., \& {Wyckoff}, S. 2005, \apj, 634, 1126

\bibitem[{{Minissale} {et~al.}(2022){Minissale}, {Aikawa}, {Bergin}, {Bertin}, {Brown}, {Cazaux}, {Charnley}, {Coutens}, {Cuppen}, {Guzman}, {Linnartz}, {McCoustra}, {Rimola}, {Schrauwen}, {Toubin}, {Ugliengo}, {Watanabe}, {Wakelam}, \& {Dulieu}}]{Minissale2022}
{Minissale}, M., {Aikawa}, Y., {Bergin}, E., {et~al.} 2022, ACS Earth and Space Chemistry, 6, 597

\bibitem[{{Moskienko} \& {Dyubko}(1991)}]{Moskienko1991}
{Moskienko}, E.~M. \& {Dyubko}, S.~F. 1991, Radiophysics and Quantum Electronics, 34, 181

\bibitem[{{M{\"u}ller} {et~al.}(2005){M{\"u}ller}, {Schl{\"o}der}, {Stutzki}, \& {Winnewisser}}]{Muller2005}
{M{\"u}ller}, H. S.~P., {Schl{\"o}der}, F., {Stutzki}, J., \& {Winnewisser}, G. 2005, Journal of Molecular Structure, 742, 215

\bibitem[{{M{\"u}ller} {et~al.}(2001){M{\"u}ller}, {Thorwirth}, {Roth}, \& {Winnewisser}}]{Muller2001}
{M{\"u}ller}, H.~S.~P., {Thorwirth}, S., {Roth}, D.~A., \& {Winnewisser}, G. 2001, \aap, 370, L49

\bibitem[{{Murillo} {et~al.}(2013){Murillo}, {Lai}, {Bruderer}, {Harsono}, \& {van Dishoeck}}]{Murillo2013}
{Murillo}, N.~M., {Lai}, S.-P., {Bruderer}, S., {Harsono}, D., \& {van Dishoeck}, E.~F. 2013, \aap, 560, A103

\bibitem[{{Nazari} {et~al.}(2022{\natexlab{a}}){Nazari}, {Meijerhof}, {van Gelder}, {Ahmadi}, {van Dishoeck}, {Tabone}, {Langeroodi}, {Ligterink}, {Jaspers}, {Beltr{\'a}n}, {Fuller}, {S{\'a}nchez-Monge}, \& {Schilke}}]{Nazari2022ALMAGAL}
{Nazari}, P., {Meijerhof}, J.~D., {van Gelder}, M.~L., {et~al.} 2022{\natexlab{a}}, \aap, 668, A109

\bibitem[{{Nazari} {et~al.}(2023{\natexlab{a}}){Nazari}, {Tabone}, \& {Rosotti}}]{Nazari2023}
{Nazari}, P., {Tabone}, B., \& {Rosotti}, G.~P. 2023{\natexlab{a}}, \aap, 671, A107

\bibitem[{{Nazari} {et~al.}(2022{\natexlab{b}}){Nazari}, {Tabone}, {Rosotti}, {van Gelder}, {Meshaka}, \& {van Dishoeck}}]{Nazari2022}
{Nazari}, P., {Tabone}, B., {Rosotti}, G.~P., {et~al.} 2022{\natexlab{b}}, \aap, 663, A58

\bibitem[{{Nazari} {et~al.}(2023{\natexlab{b}}){Nazari}, {Tabone}, {van't Hoff}, {J{\o}rgensen}, \& {van Dishoeck}}]{Nazari2023CGD}
{Nazari}, P., {Tabone}, B., {van't Hoff}, M. L.~R., {J{\o}rgensen}, J.~K., \& {van Dishoeck}, E.~F. 2023{\natexlab{b}}, \apjl, 951, L38

\bibitem[{{Nazari} {et~al.}(2021){Nazari}, {van Gelder}, {van Dishoeck}, {Tabone}, {van't Hoff}, {Ligterink}, {Beuther}, {Boogert}, {Caratti o Garatti}, {Klaassen}, {Linnartz}, {Taquet}, \& {Tychoniec}}]{Nazari2021}
{Nazari}, P., {van Gelder}, M.~L., {van Dishoeck}, E.~F., {et~al.} 2021, \aap, 650, A150

\bibitem[{{{\"O}berg} {et~al.}(2011){{\"O}berg}, {Boogert}, {Pontoppidan}, {van den Broek}, {van Dishoeck}, {Bottinelli}, {Blake}, \& {Evans}}]{Oberg2011}
{{\"O}berg}, K.~I., {Boogert}, A.~C.~A., {Pontoppidan}, K.~M., {et~al.} 2011, \apj, 740, 109

\bibitem[{{Penteado} {et~al.}(2017){Penteado}, {Walsh}, \& {Cuppen}}]{Penteado2017}
{Penteado}, E.~M., {Walsh}, C., \& {Cuppen}, H.~M. 2017, \apj, 844, 71

\bibitem[{{Qu{\'e}nard} {et~al.}(2018){Qu{\'e}nard}, {Jim{\'e}nez-Serra}, {Viti}, {Holdship}, \& {Coutens}}]{Quenard2018}
{Qu{\'e}nard}, D., {Jim{\'e}nez-Serra}, I., {Viti}, S., {Holdship}, J., \& {Coutens}, A. 2018, \mnras, 474, 2796

\bibitem[{{Rivilla} {et~al.}(2017){Rivilla}, {Beltr{\'a}n}, {Cesaroni}, {Fontani}, {Codella}, \& {Zhang}}]{Rivilla2017}
{Rivilla}, V.~M., {Beltr{\'a}n}, M.~T., {Cesaroni}, R., {et~al.} 2017, \aap, 598, A59

\bibitem[{{Sch{\"o}ier} {et~al.}(2005){Sch{\"o}ier}, {van der Tak}, {van Dishoeck}, \& {Black}}]{Schoier2005}
{Sch{\"o}ier}, F.~L., {van der Tak}, F.~F.~S., {van Dishoeck}, E.~F., \& {Black}, J.~H. 2005, \aap, 432, 369

\bibitem[{{Skouteris} {et~al.}(2018){Skouteris}, {Balucani}, {Ceccarelli}, {Vazart}, {Puzzarini}, {Barone}, {Codella}, \& {Lefloch}}]{Skouteris2018}
{Skouteris}, D., {Balucani}, N., {Ceccarelli}, C., {et~al.} 2018, \apj, 854, 135

\bibitem[{{Slavicinska} {et~al.}(2023){Slavicinska}, {Rachid}, {Rocha}, {Chuang}, {van Dishoeck}, \& {Linnartz}}]{Slavicinska2023}
{Slavicinska}, K., {Rachid}, M.~G., {Rocha}, W.~R.~M., {et~al.} 2023, \aap, 677, A13

\bibitem[{{Suzuki} {et~al.}(2018){Suzuki}, {Ohishi}, {Saito}, {Hirota}, {Majumdar}, \& {Wakelam}}]{Suzuki2018}
{Suzuki}, T., {Ohishi}, M., {Saito}, M., {et~al.} 2018, \apjs, 237, 3

\bibitem[{{Taniguchi} {et~al.}(2023){Taniguchi}, {Sanhueza}, {Olguin}, {Gorai}, {Das}, {Nakamura}, {Saito}, {Zhang}, {Lu}, {Li}, \& {Chen}}]{Taniguchi2023}
{Taniguchi}, K., {Sanhueza}, P., {Olguin}, F.~A., {et~al.} 2023, arXiv e-prints, arXiv:2304.00267

\bibitem[{{Taquet} {et~al.}(2015){Taquet}, {L{\'o}pez-Sepulcre}, {Ceccarelli}, {Neri}, {Kahane}, \& {Charnley}}]{Taquet2015}
{Taquet}, V., {L{\'o}pez-Sepulcre}, A., {Ceccarelli}, C., {et~al.} 2015, \apj, 804, 81

\bibitem[{{Tobin} {et~al.}(2020){Tobin}, {Sheehan}, {Megeath}, {D{\'\i}az-Rodr{\'\i}guez}, {Offner}, {Murillo}, {van 't Hoff}, {van Dishoeck}, {Osorio}, {Anglada}, {Furlan}, {Stutz}, {Reynolds}, {Karnath}, {Fischer}, {Persson}, {Looney}, {Li}, {Stephens}, {Chandler}, {Cox}, {Dunham}, {Tychoniec}, {Kama}, {Kratter}, {Kounkel}, {Mazur}, {Maud}, {Patel}, {Perez}, {Sadavoy}, {Segura-Cox}, {Sharma}, {Stephenson}, {Watson}, \& {Wyrowski}}]{Tobin2020}
{Tobin}, J.~J., {Sheehan}, P.~D., {Megeath}, S.~T., {et~al.} 2020, \apj, 890, 130

\bibitem[{{van der Tak} {et~al.}(2000){van der Tak}, {van Dishoeck}, {Evans}, \& {Blake}}]{vanderTak2000}
{van der Tak}, F. F.~S., {van Dishoeck}, E.~F., {Evans}, Neal~J., I., \& {Blake}, G.~A. 2000, \apj, 537, 283

\bibitem[{{van Dishoeck} {et~al.}(1995){van Dishoeck}, {Blake}, {Jansen}, \& {Groesbeck}}]{Ewine1995}
{van Dishoeck}, E.~F., {Blake}, G.~A., {Jansen}, D.~J., \& {Groesbeck}, T.~D. 1995, \apj, 447, 760

\bibitem[{{van Gelder} {et~al.}(2022{\natexlab{a}}){van Gelder}, {Jaspers}, {Nazari}, {Ahmadi}, {van Dishoeck}, {Beltr{\'a}n}, {Fuller}, {S{\'a}nchez-Monge}, \& {Schilke}}]{vanGelder2022deuteration}
{van Gelder}, M.~L., {Jaspers}, J., {Nazari}, P., {et~al.} 2022{\natexlab{a}}, \aap, 667, A136

\bibitem[{{van Gelder} {et~al.}(2022{\natexlab{b}}){van Gelder}, {Nazari}, {Tabone}, {Ahmadi}, {van Dishoeck}, {Beltr{\'a}n}, {Fuller}, {Sakai}, {S{\'a}nchez-Monge}, {Schilke}, {Yang}, \& {Zhang}}]{vanGelder2022}
{van Gelder}, M.~L., {Nazari}, P., {Tabone}, B., {et~al.} 2022{\natexlab{b}}, \aap, 662, A67

\bibitem[{{van Gelder} {et~al.}(2020){van Gelder}, {Tabone}, {Tychoniec}, {van Dishoeck}, {Beuther}, {Boogert}, {Caratti o Garatti}, {Klaassen}, {Linnartz}, {M{\"u}ller}, \& {Taquet}}]{vanGelder2020}
{van Gelder}, M.~L., {Tabone}, B., {Tychoniec}, {\L}., {et~al.} 2020, \aap, 639, A87

\bibitem[{{Vastel} {et~al.}(2015){Vastel}, {Bottinelli}, {Caux}, {Glorian}, \& {Boiziot}}]{Vastel2015}
{Vastel}, C., {Bottinelli}, S., {Caux}, E., {Glorian}, J.~M., \& {Boiziot}, M. 2015, in SF2A-2015: Proceedings of the Annual meeting of the French Society of Astronomy and Astrophysics, 313--316

\bibitem[{{Wakelam} {et~al.}(2017){Wakelam}, {Loison}, {Mereau}, \& {Ruaud}}]{Wakelam2017}
{Wakelam}, V., {Loison}, J.~C., {Mereau}, R., \& {Ruaud}, M. 2017, Molecular Astrophysics, 6, 22

\bibitem[{{Walsh} {et~al.}(2014){Walsh}, {Millar}, {Nomura}, {Herbst}, {Widicus Weaver}, {Aikawa}, {Laas}, \& {Vasyunin}}]{Walsh2014}
{Walsh}, C., {Millar}, T.~J., {Nomura}, H., {et~al.} 2014, \aap, 563, A33

\bibitem[{{Watanabe} \& {Kouchi}(2002)}]{Watanabe2002}
{Watanabe}, N. \& {Kouchi}, A. 2002, \apjl, 571, L173

\bibitem[{{Williams} {et~al.}(2022){Williams}, {Cyganowski}, {Brogan}, {Hunter}, {Ilee}, {Nazari}, {Kruijssen}, {Smith}, \& {Bonnell}}]{Williams2022}
{Williams}, G.~M., {Cyganowski}, C.~J., {Brogan}, C.~L., {et~al.} 2022, mnras, 509, 748

\bibitem[{{Wilson} \& {Rood}(1994)}]{Wilson1994}
{Wilson}, T.~L. \& {Rood}, R. 1994, \araa, 32, 191

\bibitem[{{Xu} {et~al.}(2008){Xu}, {Fisher}, {Lees}, {Shi}, {Hougen}, {Pearson}, {Drouin}, {Blake}, \& {Braakman}}]{Xu2008}
{Xu}, L.-H., {Fisher}, J., {Lees}, R.~M., {et~al.} 2008, Journal of Molecular Spectroscopy, 251, 305

\bibitem[{{Yang} {et~al.}(2021){Yang}, {Sakai}, {Zhang}, {Murillo}, {Zhang}, {Higuchi}, {Zeng}, {L{\'o}pez-Sepulcre}, {Yamamoto}, {Lefloch}, {Bouvier}, {Ceccarelli}, {Hirota}, {Imai}, {Oya}, {Sakai}, \& {Watanabe}}]{Yang2021}
{Yang}, Y.-L., {Sakai}, N., {Zhang}, Y., {et~al.} 2021, \apj, 910, 20

\end{thebibliography}

\begin{appendix}

\section{Comparison between CASSIS fits and the true column densities}
\label{sec:CASSIS_true}

Here we check how the fitted CASSIS column densities compare with the true column densities. To ease this comparison we derived the true total number of molecules in a $1\arcsec$ radius around the protostar and we then compare this value to the total number of molecules found from CASSIS in a circular beam with $1\arcsec$ radius. 

The total number of molecules from our CASSIS fits is given by $N \pi R_{\rm beam}^2$, where $N$ is the column density and $R_{\rm beam}$ is the radius of the beam. This radius is 150\,au and 4000\,au for the low- and high-mass models. The true total number of molecules in the models is calculated as 

\begin{equation}
    \mathcal{N} = \int_{0}^{r_{\rm beam}} \int_{-\pi/2}^{\pi/2} 2 \pi X(r,\theta) n_{\rm H}(r, \theta) r^2 sin(\theta) d\theta dr,
    \label{eq:true_N}
\end{equation}

\noindent where $r$ is the radius in spherical coordinates, $r_{\rm beam}$ is 150\,au and 4000\,au for the low- and high-mass models, $X$ is the abundance of the molecule in the gas phase and $n_{\rm H}$ is the hydrogen nucleus number density. 

\begin{figure}
  \resizebox{\hsize}{!}{\includegraphics{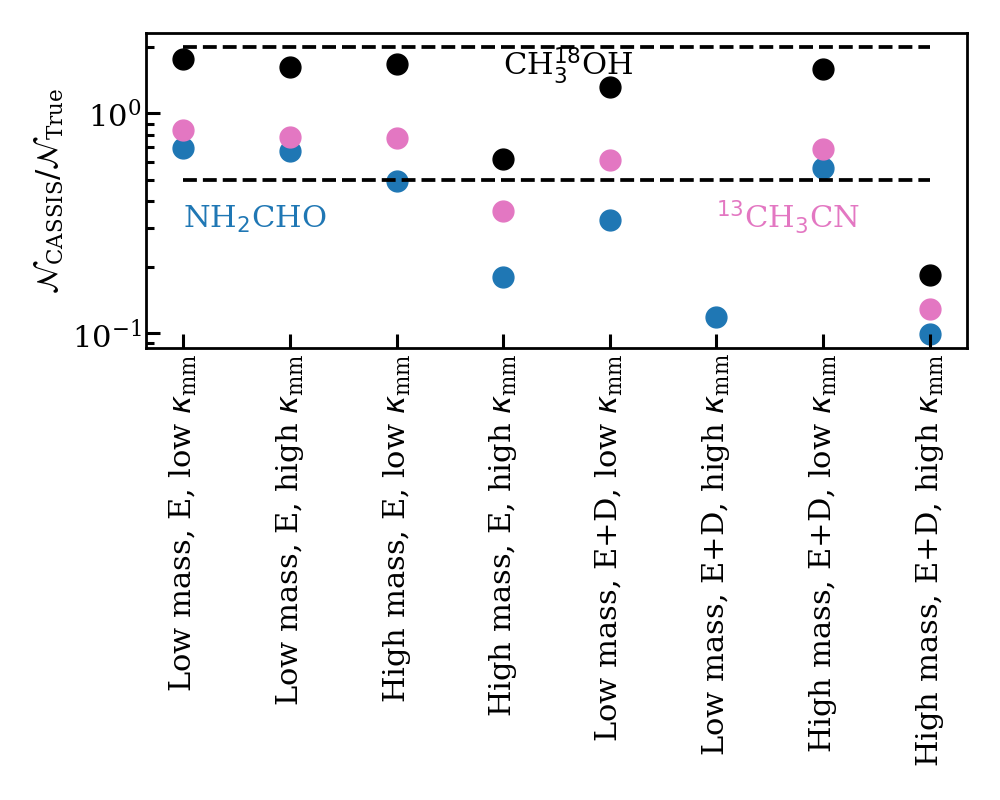}}
  \caption{Comparison of true number of NH$_2$CHO, CH$_3^{18}$OH, and $^{13}$CH$_3$CN molecules and that found from the CASSIS fits for our fiducial models and those with high mm opacity dust. For the low-mass model with disk and high mm opacity dust methanol and methyl cyanide are not detected due to dust optical depth effects. The two dashed lines highlight the region for a factor of two difference between the true value and that found from CASSIS.}
  \label{fig:true_CASSIS}
\end{figure}

Figure \ref{fig:true_CASSIS} presents the ratio of the number of NH$_2$CHO, CH$_3^{18}$OH, and $^{13}$CH$_3$CN molecules found from CASSIS fits and the true number of molecules in the fiducial models and those with high mm opacity dust grains. This shows that in general there is good agreement between these two values (within a factor of $\lesssim 2$).

For the other models with lower CASSIS values (larger difference with a factor of $> 3$) the dust has a high mm opacity. This can be explained by dust opacity effects. This is particularly noticeable for the low-mass model with a disk and high mm opacity dust where methanol and methyl cyanide are not detected at all and formamide has one order of magnitude lower value found from CASSIS while the true numbers of molecules within the $1\arcsec$ radius for the three molecules are similar (within ${\sim}50\%$) to the low-mass model with disk and low mm opacity dust. These results agree well with the conclusions of \citet{Nazari2022,Nazari2023} to explain the low methanol emission from some protostellar systems. It is interesting to note that for the envelope-only low-mass protostar with high mm opacity dust the CASSIS values are not underestimated which is expected from the results of \cite{Nazari2022} where they concluded that optically thick dust alone cannot explain the low methanol emission around protostars.
\section{Additional tables and plots}
\label{sec:add_plots}

\begin{figure*}
    \centering
    \includegraphics[width=0.8\textwidth]{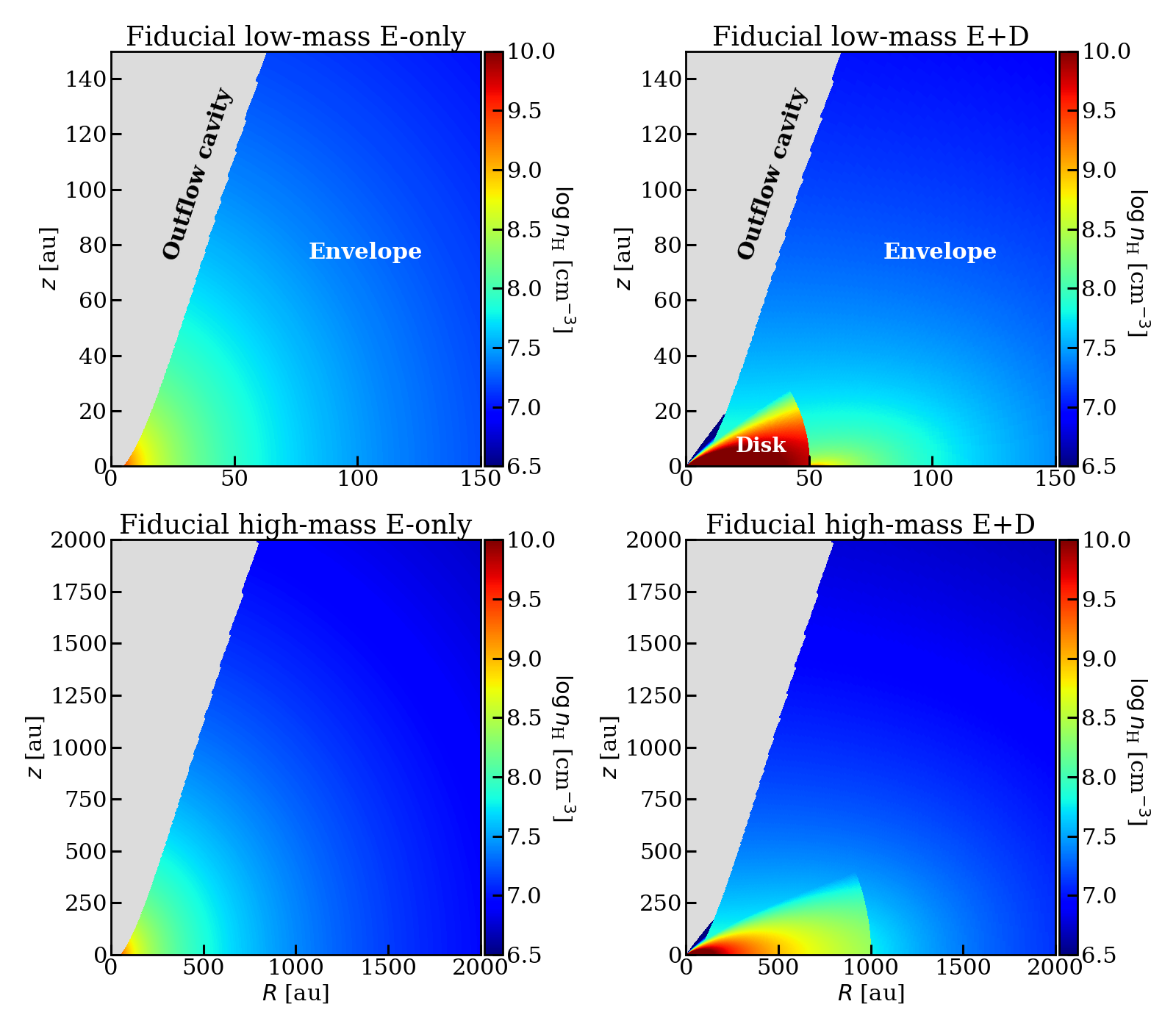}
    \caption{Hydrogen nuclear number density of the fiducial low- and high-mass models. The left column shows the envelope-only models, while the right column shows the envelope-plus-disk models.} 
    \label{fig:density_scatterpaper}
\end{figure*}

\begin{figure*}
    \centering
    \includegraphics[width=0.8\textwidth]{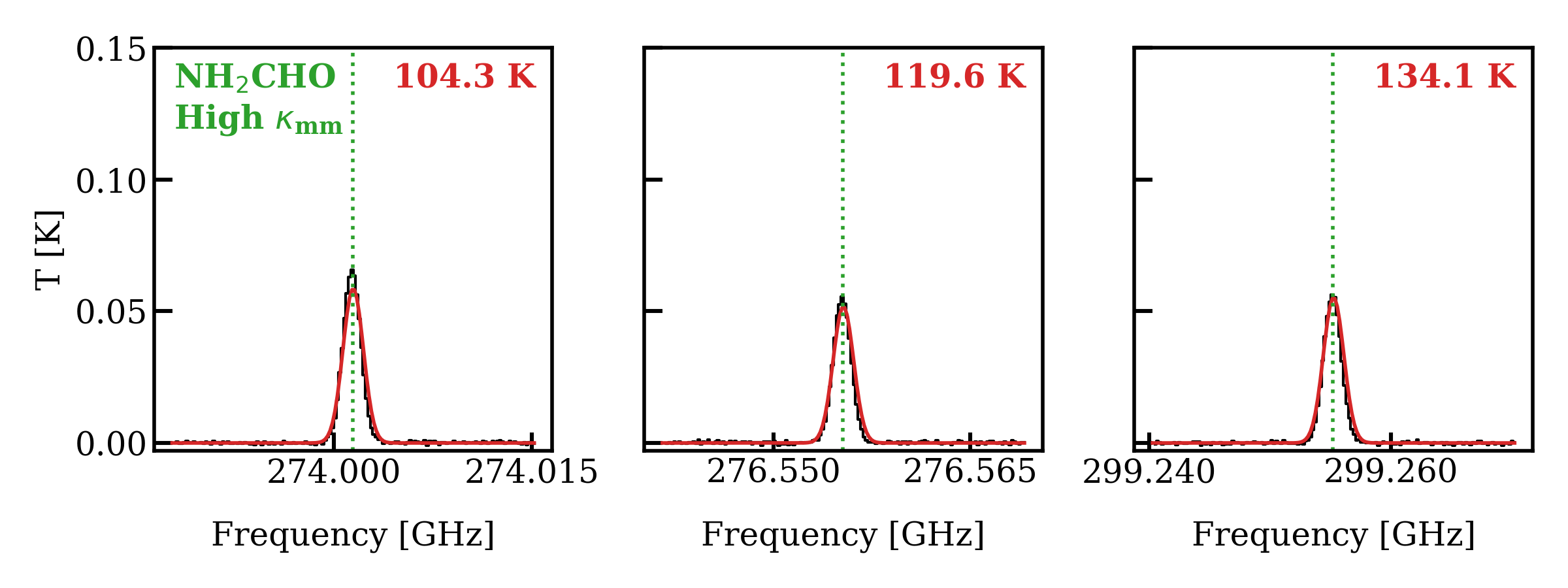}
    \caption{Same as Fig. \ref{fig:fit_fid} but for the fiducial low-mass model with high mm opacity dust (i.e., large dust grains). The excitation temperature is fixed to 150\,K.} 
    \label{fig:fit_fid_large}
\end{figure*}

\begin{figure*}
    \centering
    \includegraphics[width=0.8\textwidth]{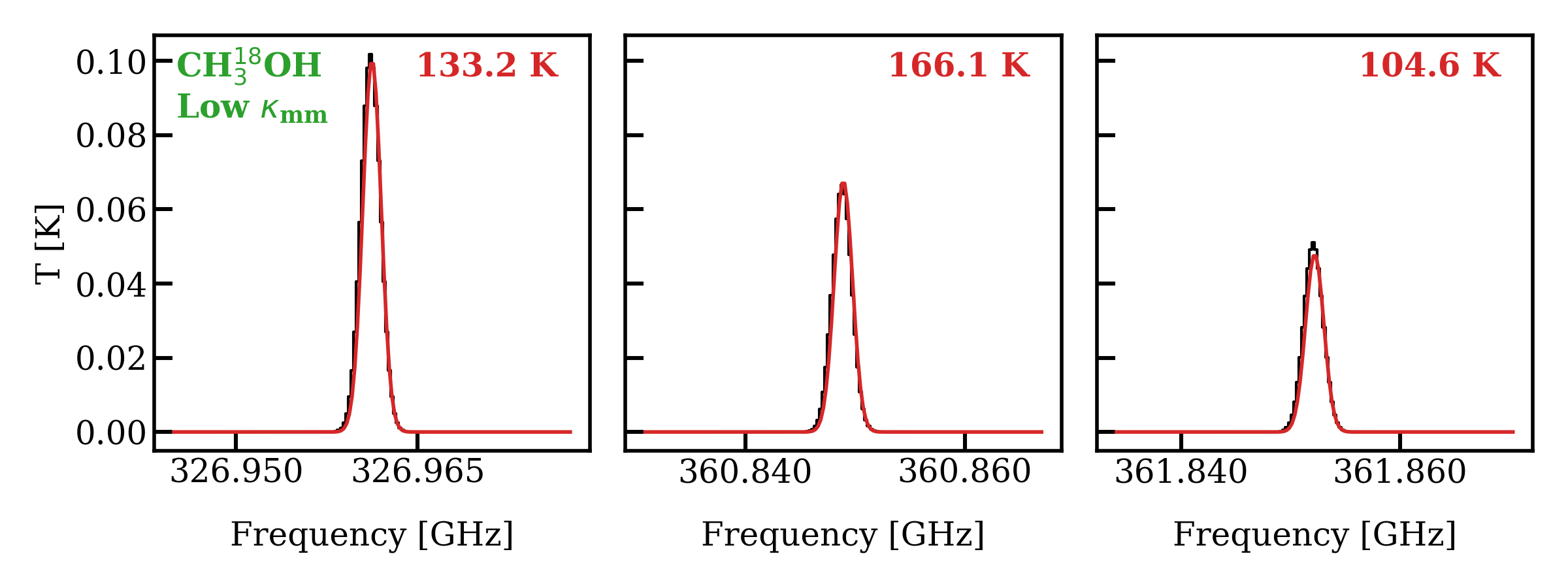}
    \caption{Same as Fig. \ref{fig:fit_fid} but for the fiducial low-mass model of CH$_3^{18}$OH. The excitation temperature is fixed to 150\,K.} 
    \label{fig:fit_fid_CH3OH}
\end{figure*}

\begin{figure*}
    \centering
    \includegraphics[width=0.8\textwidth]{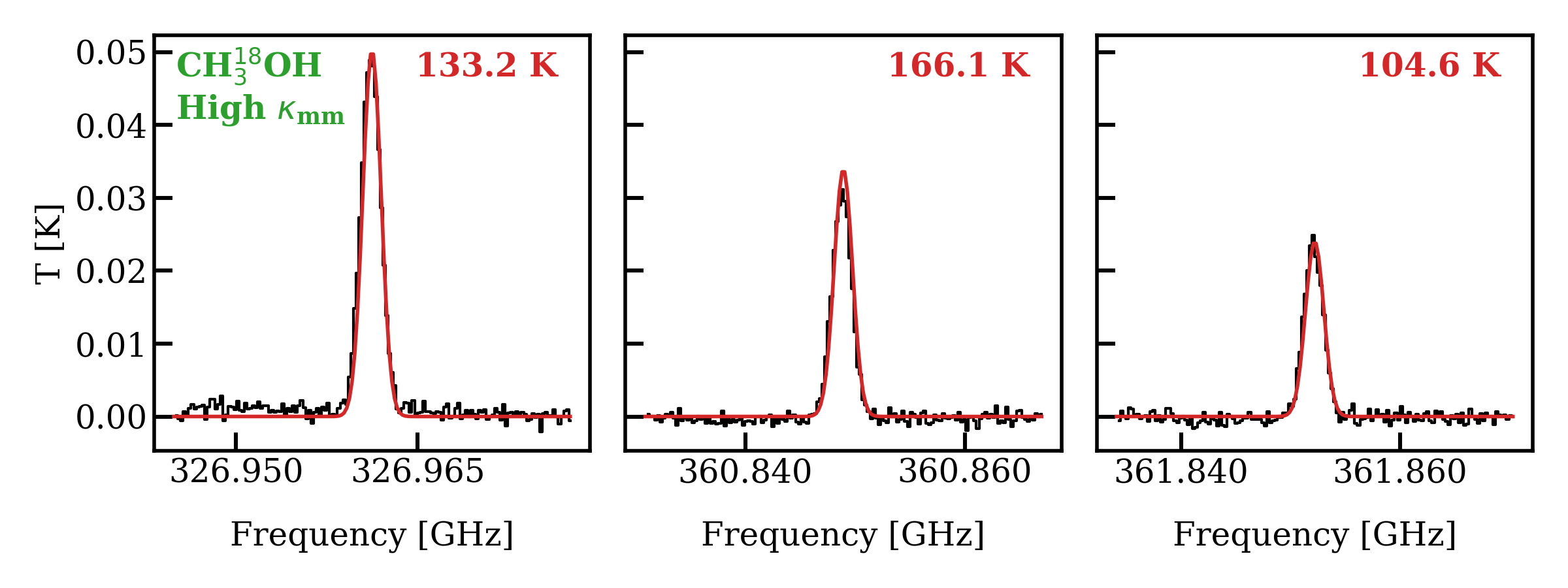}
    \caption{Same as Fig. \ref{fig:fit_fid} but for the fiducial low-mass model of CH$_3^{18}$OH with high mm opacity dust (i.e., large dust grains). The excitation temperature is fixed to 150\,K.} 
    \label{fig:fit_fid_large_CH3OH}
\end{figure*}

\begin{figure*}
    \centering
    \includegraphics[width=0.8\textwidth]{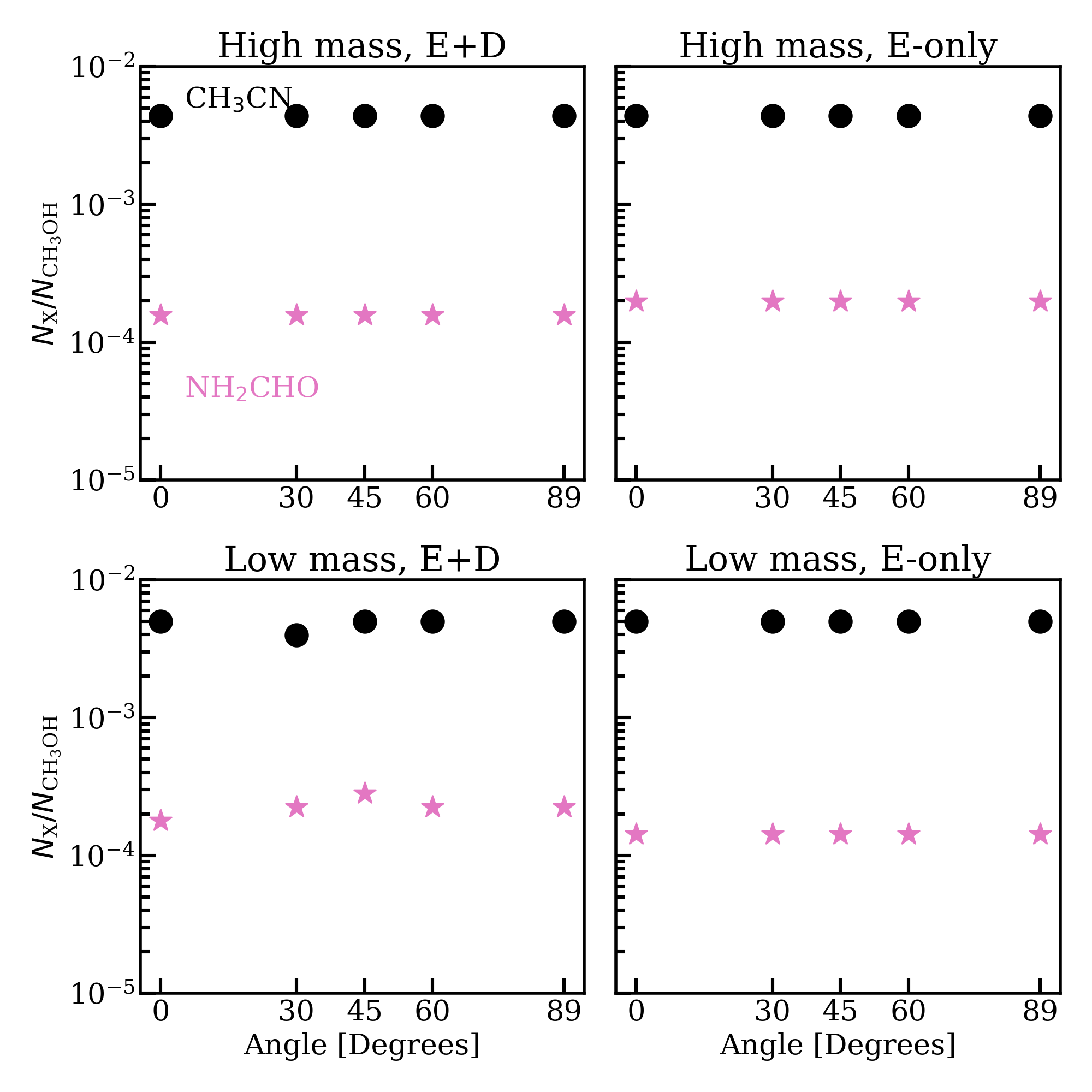}
    \caption{Column density ratios of formamide (pink stars) and methyl cyanide (black circles) to methanol for the fiducial models but calculated with the lines ray traced with different viewing angles.} 
    \label{fig:angles}
\end{figure*}

\begin{figure*}
    \centering
    \includegraphics[width=0.8\textwidth]{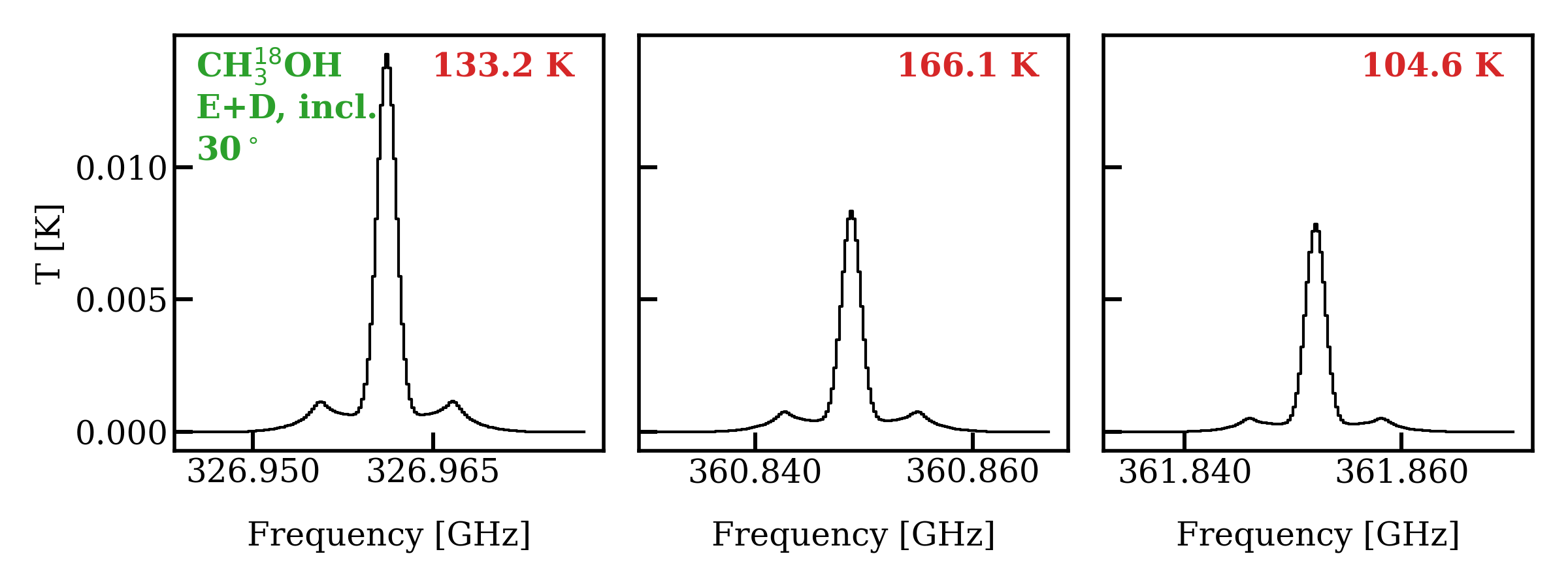}
    \caption{Lines of CH$_3^{18}$OH for the low-mass envelope-plus-disk fiducial model with viewing angle of 30$^\circ$ to show case the triply peaked profile.} 
    \label{fig:lines_30_methanol}
\end{figure*}

\begin{figure*}
    \centering
    \includegraphics[width=0.8\textwidth]{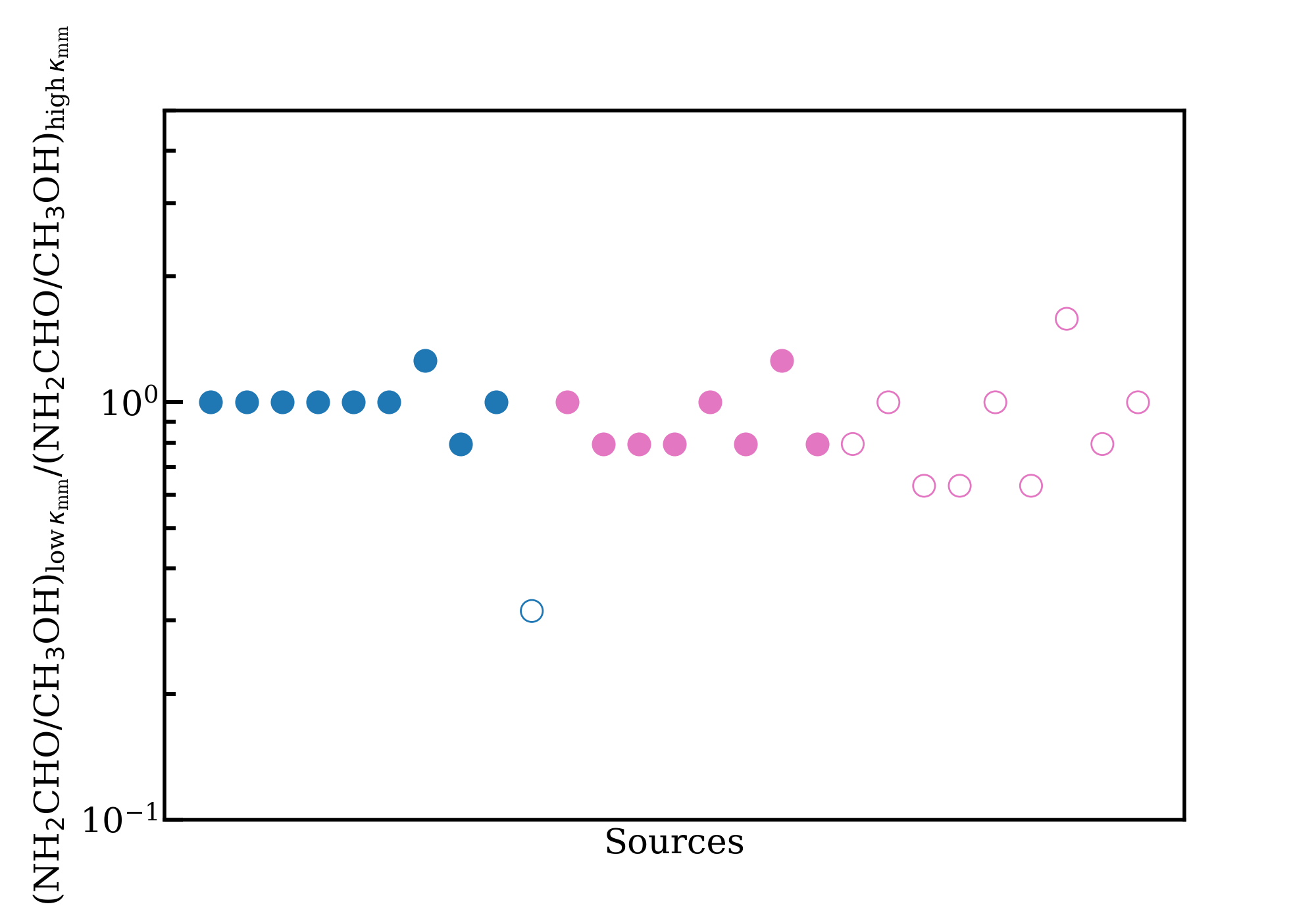}
    \caption{Ratios of $(\frac{\rm{NH}_2\rm{CHO}}{\rm{CH}_3\rm{OH}})_{\rm{low}\,\kappa_{\rm mm}}/(\frac{\rm{NH}_2\rm{CHO}}{\rm{CH}_3\rm{OH}})_{\rm{high}\,\kappa_{\rm mm}}$, where the numerator and denominator correspond to the same models with the only difference being low mm opacity versus high mm opacity dust in the models. Pink shows the high-mass models and blue the low-mass ones. Filled circles are the envelope-only models and the empty circles are those with a disk. Methanol in most of the low-mass envelope-plus-disk models with optically thick dust is not detected and thus there is only one blue empty circle. The effect of dust, alone, is less than a factor of 3, thus most of the scatter in Fig. \ref{fig:combined} is driven by the difference in source structure.} 
    \label{fig:dust}
\end{figure*}

\begin{figure*}
    \centering
    \includegraphics[width=0.8\textwidth]{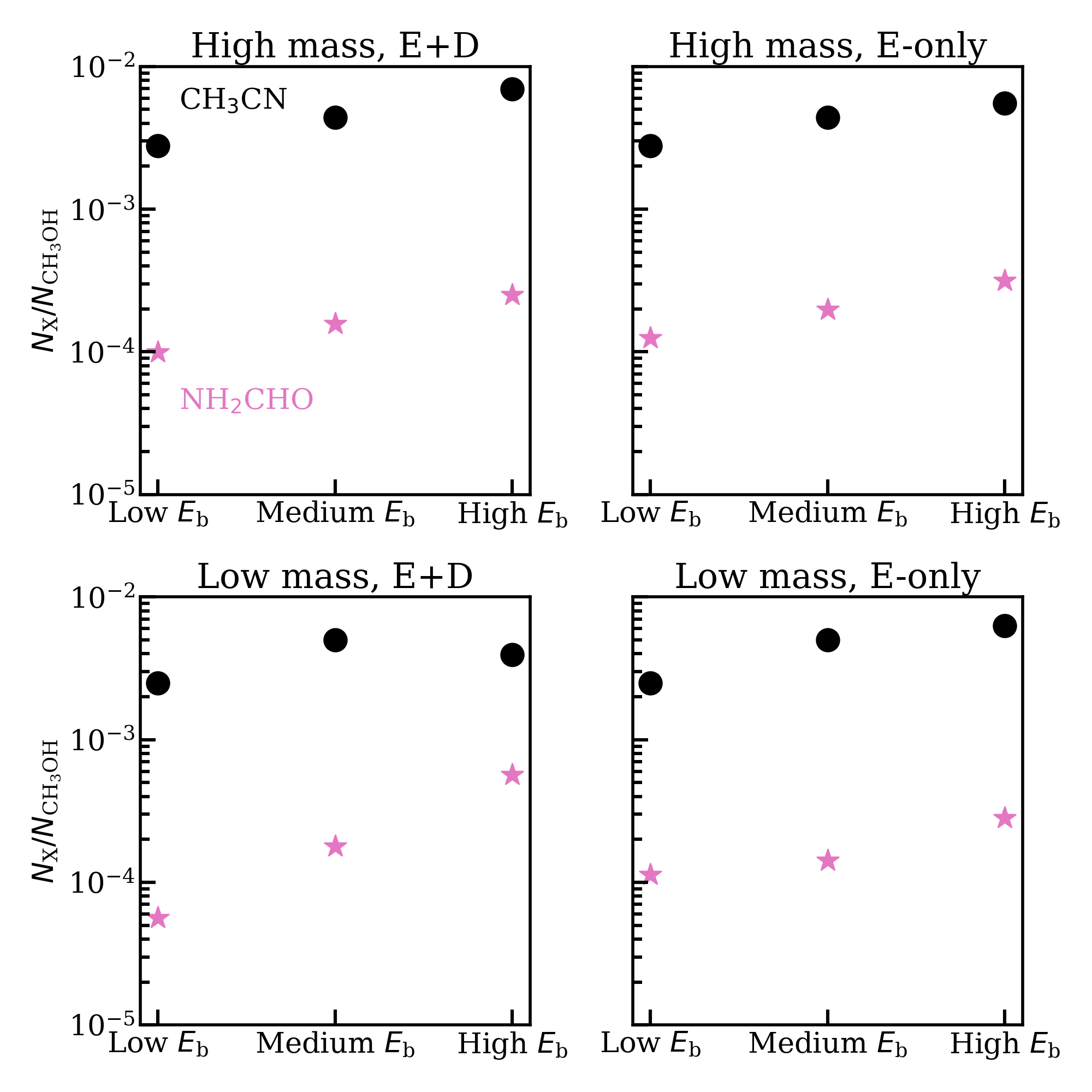}
    \caption{Column density ratios of formamide (pink stars) and methyl cyanide (black circles) to methanol for the fiducial models but varying the binding energies of each molecule based on the range of values reported in the literature (\citealt{Penteado2017}; \citealt{Ferrero2020}; \citealt{Busch2022}; \citealt{Minissale2022}). `Low $E_{\rm b}$' corresponds to 5500\,K, 3500\,K, and 3500\,K for formamide, methanol and methyl cyanide. `High $E_{\rm b}$' corresponds to 1100\,K, 8600\,K, and 7600\,K for these species. `Medium $E_{\rm b}$' corresponds to the fiducial models which have binding energies in between of the range. For all of these models the same pre-factor as the fiducial models was assumed when calculating the ice and gas abundances.} 
    \label{fig:min_max}
\end{figure*}

Table \ref{tab:lines} presents the spectral lines that are considered for each molecule in this work. Figure \ref{fig:density_scatterpaper} presents the density structures for the fiducial low- and high-mass models. Figures \ref{fig:fit_fid_large}-\ref{fig:fit_fid_large_CH3OH} show the fitted CASSIS models on top of the results from the fiducial low-mass models and those with high mm opacity dust for formamide and methanol lines. Figure \ref{fig:angles} presents the column density ratios of formamide and methyl cyanide to methanol in the fiducial models with varying viewing angles. Figure \ref{fig:lines_30_methanol} presents the line profiles of $^{18}$O methanol for the fiducial low-mass envelope-plus-disk model with an inclination angle of 30$^{\circ}$. Figure \ref{fig:dust} presents how much scatter in NH$_2$CHO/CH$_3$OH is produced by the dust alone. Figure \ref{fig:min_max} illustrates the column density ratios of NH$_2$CHO/CH$_3$OH and CH$_3$CN/CH$_3$OH for the fiducial models but varying the binding energies to include a range of reported values in the literature. This figure further supports the fact that a large scatter in column density ratios could be due to vastly different binding energies.

\begin{table*}
    \caption{Modeled spectral lines}
    \label{tab:lines}
    \centering
    \begin{tabular}{l l l l l} 
    \toprule
    \toprule    
Species & Transition & Frequency & $A_{\rm ij}$ & $E_{\rm up}$ \\
        & J K L (M) & (GHz) & (s$^{-1}$) & (K)\\
\midrule     

CH$_3$OH & 13 6 8 1 - 14 5 9 1 & 213.3775$^{\star}$ & $1.1 \times 10^{-5}$ & 389.9 \\
 & 4 2 3 1-3 1 2 1 &  218.4401$^{\star}$ & $4.7 \times 10^{-5}$ & 45.5 \\
 & 10 1 10 2 - 9 0 9 1 & 326.9612 & $1.3 \times 10^{-4}$ & 133.1\\
 & 13 4 9 1 - 14 3 12 1 & 327.4868$^{\star}$	& $5.6 \times 10^{-5}$ & 307.2\\
 & 11 0 11 1 - 10 1 9 1 & 360.8489 & $1.2 \times 10^{-4}$ & 166.0\\
 & 8 1 7 1 - 7 2 6 1 & 361.8522 & $7.7 \times 10^{-5}$ & 104.6\\
 & 9 8 1 1-8 8 0 1 & 434.9518$^{\star}$	& $7.8 \times 10^{-5}$ & 439.5\\
 & 10 9 2 2 - 9 9 1 2 & 483.0728$^{\star}$	& $9.7 \times 10^{-5}$ & 530.4\\
 & 11 10 1 2 - 10 10 0 2 & 531.2772$^{\star}$	& $1.2 \times 10^{-4}$ & 662.9\\
\hline
NH$_2$CHO & 10 1 9 - 9 1 8 & 218.4592$^{\star}$ & $7.5 \times 10^{-4}$	& 60.8\\
 & 12 1 12 - 11 1 11 & 243.5210$^{\star}$ & $1.1 \times 10^{-3}$ & 79.2\\
 & 13 2 12 - 12 2 11 & 274.0014 & $1.5 \times 10^{-3}$ & 104.3\\
 & 13 7 6 - 12 7 5 & 275.9945$^{\star}$ & $1.1 \times 10^{-3}$ & 238.6\\
 & 13 3 11 - 12 3 10 & 276.5553 & $1.5 \times 10^{-3}$ & 119.6\\
 & 14 3 11 - 13 3 10 & 299.2552 & $1.9 \times 10^{-3}$ & 134.1\\
 & 15 11 4 - 14 11 3 & 318.4563$^{\star}$ & $1.1 \times 10^{-3}$ & 482.0\\
 & 15 8 7 - 14 8 6 & 318.4626$^{\star}$ & $1.7 \times 10^{-3}$ & 312.7\\
 & 16 13 3 - 15 13 2 & 339.7463$^{\star}$ & $9.8 \times 10^{-4}$ &	640.5\\
 & 17 12 5 - 16 12 4 & 360.9465$^{\star}$ & $1.7 \times 10^{-3}$ & 583.8\\
 & 17 14 3 - 16 14 2 & 361.0199$^{\star}$ & $1.1 \times 10^{-3}$ &	737.6\\
\hline
CH$_3$CN & 12 3 0 - 11 -3 0 & 220.7089 & $8.4 \times 10^{-4}$ & 133.2\\
 & 12 1 0 - 11 1 0 & 220.7430$^{\star}$	& $8.4 \times 10^{-4}$ & 76.0\\
 & 12 0 0 - 11 0 0 & 220.7473$^{\star}$ & $9.2 \times 10^{-4}$ & 68.87\\
 & 15 9 0 - 14 -9 0 & 275.4845$^{\star}$	& $1.1 \times 10^{-3}$ & 683.7\\
 & 15 8 0 - 14 8 0 & 275.5741$^{\star}$ & $1.2 \times 10^{-3}$ & 562.6\\
 & 15 7 0 - 14 7 0 & 275.6548$^{\star}$ & $1.3 \times 10^{-3}$ & 455.7\\
 & 15 6 0 - 14 -6 0 & 275.7240$^{\star}$ & $1.4 \times 10^{-3}$ & 363.0\\
 & 15 5 0 - 14 5 0 & 275.7825$^{\star}$ & $1.6 \times 10^{-3}$ & 284.5\\
 & 15 4 0 - 14 4 0 & 275.8304$^{\star}$ & $1.6 \times 10^{-3}$ & 220.2\\
 & 15 3 0 - 14 -3 0 & 275.8677 & $1.7 \times 10^{-3}$ & 170.2\\
 & 15 2 0 - 14 2 0 & 275.8943$^{\star}$ & $1.7 \times 10^{-3}$ & 134.5\\
 & 15 1 0 - 14 1 0 & 275.9103$^{\star}$ & $1.7 \times 10^{-3}$ & 113.1\\
 & 15 0 0 - 14 0 0 & 275.9156$^{\star}$ & $1.8 \times 10^{-3}$ & 105.9\\
 & 16 3 0 - 15 -3 0 & 294.2513 & $2.1 \times 10^{-3}$ & 184.4\\

\bottomrule
\end{tabular}
\tablefoot{Stars indicate the lines that were only produced for the fiducial models and are used for column density measurement with varying excitation temperatures. To avoid confusion, this table only shows one line to represent all the lines that are a result of (hyper)fine splitting, this is particularly important for the chosen CH$_3$CN lines.}
\end{table*}


\end{appendix}

\end{document}